\documentclass[fleqn,usenatbib]{mnras}

\usepackage{newtxtext,newtxmath}
\usepackage[T1]{fontenc}

\usepackage{graphicx}
\usepackage{amsmath}
\usepackage{lastpage}

\DeclareRobustCommand{\VAN}[3]{#2}
\let\VANthebibliography\thebibliography
\def\thebibliography{%
  \DeclareRobustCommand{\VAN}[3]{##3}%
  \VANthebibliography
}

\title[Broad-line ambiguity in IC~1495]
{Broad-line ambiguity in the isolated Seyfert galaxy IC 1495} 
\author[I.O. Izviekova et al.]{
I.O. Izviekova,$^{1,2}$\thanks{E-mail: izviekova@gmail.com}
O.V. Kompaniiets,$^{1}$
I.B. Vavilova,$^{1}$
D.V. Dobrycheva$^{1}$
\\
$^{1}$Main Astronomical Observatory of the NAS of Ukraine, 27, Akademik Zabolotny St., Kyiv, 03143, Ukraine\\
$^{2}$International Centre for Astronomical, Medical and Ecological Research (ICAMER), 27, Akademik Zabolotny St., Kyiv, 03143, Ukraine\\
}

\date{Accepted XXX. Received November 14}

\pubyear{2026}

\begin{document}

\raggedbottom
\label{firstpage}
\pagerange{\pageref{firstpage}--\pageref{LastPage}}
\maketitle

\begin{abstract}
Published analyses disagree on the presence of broad H$\alpha$ in the
2004 6dFGS spectrum of IC~1495, a barred Seyfert in the 2MIG
catalogue of isolated galaxies. We compare this spectrum with a 2024 SDSS-V/BOSS observation
to assess whether broad-H$\alpha$ visibility changed between
the two epochs. The 2024 [O~III] profile favours a
blueshifted wing with intrinsic full width at half-maximum (FWHM)
of $\simeq640~\mathrm{km\,s^{-1}}$. Single-component fits are
preferred for [S~II] and [O~I]. Among six 2024 H$\alpha$+[N~II]
fits with free narrow-line kinematics, the fit preferred by the
Bayesian information criterion (BIC) includes broad H$\alpha$
with intrinsic FWHM of $\simeq2000~\mathrm{km\,s^{-1}}$.
Adding broad H$\alpha$ increases the BIC when H$\alpha$ and [N~II]
share two narrow kinematic components. Broad H$\alpha$ is
undetected in 2004 under the adopted criterion. A reference broad
component with intrinsic FWHM $\simeq1930~\mathrm{km\,s^{-1}}$
and a broad-to-narrow H$\alpha$ flux ratio of 0.82 is recovered
in 27--35 per cent of baseline simulations and 11--55 per cent
across the grid. At 1.5 times that ratio, baseline recovery is
42--55 per cent. Aperture and flux-calibration uncertainties
limit absolute line-flux comparisons. Narrow-line excitation,
\textit{Chandra} spectroscopy
($N_{\rm H,int}\simeq2.3\times10^{23}~\mathrm{cm^{-2}}$ in the
fiducial model), and long-term optical and mid-infrared variability
support an absorbed, variable Seyfert nucleus. The model-dependent 2024 broad-H$\alpha$ candidate and the limited 2004 sensitivity do not establish a change
in broad-line visibility between the two epochs.
\end{abstract}


\section{Introduction}
\label{sec:introduction}

Weak broad Balmer emission can affect the optical type assigned
to a Seyfert nucleus. Its visibility depends on the intrinsic
strength of emission from the broad-line region (BLR), attenuation
towards the emitting gas, and contrast against the stellar
continuum and narrow lines in the spectroscopic aperture
\citep{Trippe2010}. X-ray spectroscopy probes the absorbing gas
column towards the compact X-ray source, which can differ from the
column towards the optically visible BLR \citep{Shimizu2018BASSVIII}.

The 2MASS-selected Isolated Galaxies catalogue (2MIG) provides
the parent sample for our multiwavelength studies of active
galactic nuclei (AGNs) and their hosts
\citep{Karachentseva2010,Pulatova2015}. X-ray spectroscopy has
identified heavily absorbed Seyfert nuclei in this sample
\citep{Kompaniiets2026_2MIG}, motivating an examination of their
optical broad-line visibility.
We begin with IC~1495
(2MIG~3166), a nearby barred galaxy previously classified as
Seyfert~2 \citep{Maia2003}. \citet{Senarath2021} included IC~1495 among
WISE-variable type-2 AGNs considered in their
search for changing-look sources. Two analyses of its 2004 Six-degree
Field Galaxy Survey (6dFGS) spectrum assigned different types.
\citet{Chen2022} reported Seyfert~1 with broad H$\alpha$, and
\citet{Suresh2025} classified the same observation as Type~2.
These classifications motivate a reassessment of which spectral
components the 2004 data support. An independent SDSS-V/BOSS
spectrum obtained in 2024 extends the comparison over nearly
two decades.

The identification of weak broad Balmer emission in host-dominated
spectra depends on the decomposition of stellar and nuclear light
\citep{Ren2024,Aydar2026Decomposition}. The shape of the narrow
lines is especially important in the H$\alpha$+[N~II] blend.
Asymmetric or multiple narrow components can leave residuals
that a simpler model assigns to broad H$\alpha$
\citep{BalmaverdeCapetti2014,Shimizu2018BASSVIII}. Applying a
forbidden-line template to the blend assumes that its narrow
components sample gas with similar line-of-sight velocity
distributions. The strengths and widths of line wings can
vary among transitions \citep{KovacevicDojcinovic2022}.
Independent fits to forbidden lines can therefore identify
kinematic structure in the narrow-line region (NLR) and
motivate alternative fits to the Balmer blend.

A Seyfert~1.9 classification combines a detection of broad
H$\alpha$ with a non-detection of broad H$\beta$. Dust
attenuation, intrinsically weak broad emission, and limited
sensitivity to H$\beta$ can all produce this appearance
\citep{Trippe2010}. The broad Balmer decrement depends on dust,
line optical depths, collisional excitation, and line
responsivity \citep{Gaskell2023Reddening,Son2025Balmer}.
The interpretation of an H$\alpha$-only broad-line spectrum
therefore depends on the sensitivity to H$\beta$.

Changes in broad Balmer emission can alter the optical Seyfert
type \citep{MacLeod2016,Runnoe2016,Trakhtenbrot2019}. Such
changing-look transitions can involve changes in the ionising
continuum or in line-of-sight obscuration \citep{Ricci2023}.
Comparisons across epochs are also sensitive to differences
in spectral resolution, noise, and flux calibration
\citep{Hon2025,Amrutha2026}. For example, \citet{Amrutha2026}
matched the aperture and resolution of follow-up spectra to
those of 6dFGS to reduce instrumental differences. For IC~1495,
the two fibre apertures can admit different amounts of stellar
light and extended narrow-line emission relative to the compact
BLR. The ability to recover weak broad emission in each
observation is therefore central to assessing a change.

When the broad component is constrained to have non-negative
flux, the null value lies at the boundary of the parameter
space and its centroid and width are unidentified under the
null. The usual asymptotic calibration of a likelihood-ratio
test can then fail \citep{Protassov2002}. Simulations of
narrow-only spectra can estimate the frequency of false
detections for a given search. Simulations that include a
broad line of prescribed width and strength estimate the
probability of recovering that signal \citep{Kashyap2010}.

We ask whether either spectrum of IC~1495 requires a BLR
contribution and what the observations establish about a change
in broad H$\alpha$. We fit the 2024 [O~III], [S~II], and [O~I]
profiles independently of the H$\alpha$+[N~II] blend and test the
support for broad H$\alpha$ across continuum models and narrow-line
prescriptions. We assess the sensitivity to broad H$\beta$
using the intrinsic width and centroid of the H$\alpha$ reference
fit. We repeat the search on simulated spectra sampled on the
native 6dFGS grid, using trials with zero broad flux to assess
false detections. At the intrinsic width of the 2024 reference
component, we test three non-zero broad-to-narrow H$\alpha$
flux ratios in the baseline configuration and the reference
ratio across all eight configurations. Recovery fractions
estimate how often each injected signal would be detected
under the tested models and noise prescriptions.

\textit{Chandra} spectroscopy constrains nuclear X-ray absorption
in IC~1495, while optical and MIR monitoring characterise
aperture-integrated variability. Narrow-line diagnostics constrain
gas excitation and density, and the photometrically anchored
narrow Balmer decrement probes NLR reddening. Optical imaging
and the global H\,I profile characterise the host's stellar
structure and atomic-gas reservoir. Applying the same profile
comparisons and recovery tests to other absorbed 2MIG nuclei
would allow their optical classifications to be compared at
quantified broad-line sensitivities.

\section{IC~1495 and observational data}
\label{sec:data}

\subsection{Target and optical spectroscopy}
\label{subsec:target}

IC~1495 (IC~5327, MCG--02--59--024, PGC~71631,
IRAS~23281--1345) is a nearby barred spiral galaxy
identified as 2MIG~3166 \citep{Karachentseva2010}
and listed as SAB(r)b with a Sy2 nucleus by
\citet{Pulatova2015}.
Its coordinates and principal properties are given
in Table~\ref{tab:ic1495_properties}.
We adopt $z_{\rm sys}=0.021295\pm0.000030$ from
the Nançay H\,I heliocentric velocity
$V_{\rm hel}=6384\pm9$~km~s$^{-1}$
\citep{Theureau1998HI}, using $z=V_{\rm hel}/c$.
Distance-dependent quantities use $D_L=92.23$~Mpc.
The distance convention and angular scale are specified
in Appendix~\ref{app:data_validation}.

The historical S2 classification \citep{Maia2003}
and mixed Sy1/Sy2 literature flags in LASr
\citep{Asmus2020} provide the context for the
classification dispute described in the Introduction.
\citet{Senarath2021} reported WISE variability
of $\Delta W1>0.3$~mag and a narrow-line appearance
in archival spectroscopy obtained after 2017.
We could not unambiguously identify that spectral
product and therefore retain it only as literature
context.
The TELPERION [O~III] survey reported no distant
extended emission-line region around IC~1495
within its sensitivity, primarily probing projected
distances beyond 10~kpc \citep{Keel2022}.

Our analysis uses two independent spectroscopic
epochs separated by approximately 19.4~yr.
The 6dFGS spectrum g2330477--132908
(\texttt{SpecID}=132563), obtained on
2004 August 23 \citep{Jones2009}, covers the
principal Balmer and narrow diagnostic lines.
Its visual and red arms are treated separately
because of their different responses and resolutions.
SkyMapper photometry provides the external continuum
reference for the narrow-line flux reconstruction
and Balmer decrement.

The SDSS-V/BOSS spectrum was obtained on
2024 January 14 (MJD~60323,
\texttt{sdss\_id}=69674331)
\citep{Kollmeier2026,SDSSDR20_2026,Smee2013}.
Its source position lies approximately
$0.30^{\prime\prime}$ from the adopted optical nucleus.
The 6dFGS and BOSS fibre diameters are
$6.7^{\prime\prime}$ and $2.0^{\prime\prime}$,
respectively.
Both spectra are analysed independently on their
native grids with their instrumental resolutions.
A model-derived 2024 broad-H$\alpha$ profile
provides the reference for conditional
injection--recovery tests in the 2004 spectrum.
Aperture-transfer assumptions are assessed separately.

The optical analysis is described in
Sections~\ref{sec:optical_methods}
and~\ref{sec:broadline_methods}.
Product validation, instrumental metadata,
and the calibration-flag audit are documented in
Appendices~\ref{app:6df_data}
and~\ref{app:sdssv_data}.
The SkyMapper reference photometry and Legacy Survey imaging
are described in Appendices~\ref{app:skymapper_data}
and~\ref{app:legacy_data}, respectively.

\begin{table}
\centering
\caption{Basic properties of IC~1495.}
\label{tab:ic1495_properties}

\footnotesize
\setlength{\tabcolsep}{5pt}
\renewcommand{\arraystretch}{1.12}

\begin{tabular}{ll}
\hline
Property & Value \\
\hline
2MIG identifier & 3166 \\
RA (J2000) & $23^{\rm h}30^{\rm m}47\fs73$ \\
Dec. (J2000) & $-13^{\circ}29^{\prime}07\farcs6$ \\
Adopted systemic redshift & $0.021295\pm0.000030$ \\
Catalogue linewidth $W_{mx}$ & $304\pm18~{\rm km\,s^{-1}}$ \\
Catalogue integrated H\,I flux & $3.62~{\rm Jy\,km\,s^{-1}}$ \\
Morphology & SAB(r)b \\
Historical SSRS2 type & S2 \\
6dFGS literature classifications & Sy1 / Type~2 \\
$\log L^{\rm nuc}_{12\mu{\rm m}}$ (erg s$^{-1}$) & $42.97\pm0.22$ \\
\hline
\end{tabular}

\par\smallskip
\begin{minipage}{\columnwidth}
\footnotesize
\raggedright
\textit{Notes.}
The 2MIG identification and morphology are from
\citet{Karachentseva2010} and \citet{Pulatova2015}.
Optical classifications are from \citet{Maia2003},
\citet{Chen2022}, and \citet{Suresh2025}.
The estimated nuclear 12-$\mu$m luminosity is from
\citet{Asmus2020}, who inferred it from WISE W3 and optical
classification at the catalogue distance of 95.3~Mpc.
The systemic velocity and uncertainty used to derive the redshift
are from \citet{Theureau1998HI}. The catalogue H\,I linewidth
and flux are from EDD \citep{Courtois2009}.
\end{minipage}
\end{table}

\subsection{Photometric monitoring and X-ray observations}
\label{subsec:timedomain_xray_data}

Long-term aperture-integrated variability is studied
with ZTF $g$, $r$, and $i$ photometry
\citep{Bellm2019,Masci2019} and WISE/NEOWISE
$W1$ and $W2$ measurements
\citep{Wright2010,Mainzer2014}.
Dedicated intraday monitoring was obtained with
the 82-cm IAC80 telescope and CAMELOT2 camera
on 2025 October 13 and 14 in $R$ and $V$,
respectively.
Photometric selection and reduction are described
in Appendices~\ref{app:variability_data}
and~\ref{app:iac80_data}.
The variability analysis is described in
Section~\ref{sec:variability_analysis}.

The \textit{Chandra}/ACIS-S observation obtained
on 2020 December 5 (ObsID~23699, approximately
3.0~ks) provides nuclear imaging and a low-count
0.5--7.0~keV spectrum.
Processing and spectral analysis are described
in Appendix~\ref{app:chandra_data}
and Section~\ref{sec:xray_spectral_methods}.

\subsection{Host-galaxy photometry, imaging, and H\,I data}
\label{subsec:host_data}
\label{subsec:optical_imaging_data}
\label{subsec:hi_data}

SkyMapper DR4 $g$, $r$, and $i$ photometry
\citep{Onken2024} supplies local continuum estimates
that are combined with emission-line equivalent
widths to reconstruct the 6dFGS narrow-line fluxes.
The fiducial 6-arcsec-diameter aperture approximates
the 6.7-arcsec 6dFGS fibre, with alternative apertures
used to assess sensitivity to the mismatch.
The non-simultaneous photometry and spectroscopy
remain subject to seeing differences and source
variability.
The reconstruction is described in
Section~\ref{sec:optical_methods}.

DESI Legacy Imaging Surveys DR10 $g$, $r$, and $z$
images \citep{Dey2019} provide integrated photometry
and stellar-structure measurements.
The $r$ and $z$ bands are used for the principal
deprojected analysis, and $g$ tests its wavelength
dependence.
The structural analysis is described in
Section~\ref{sec:host_methods}.

The unresolved Nançay H\,I 21-cm profile is obtained
from the All Digital H\,I Profile Catalog in the
Extragalactic Distance Database
\citep[EDD;][]{Courtois2009}.
We analyse its integrated flux, asymmetry,
and linewidth to derive the atomic-gas mass
and linewidth-based quantities
(Section~\ref{sec:hi_methods}).
Provenance and beam-confusion tests are documented
in Appendix~\ref{app:hi_confusion}.
The profile provides no spatial constraints
on H\,I morphology, warps, or the rotation curve.

\section{Methods}
\label{sec:methods}

The optical analysis separates broad-line identification
within each epoch from cross-epoch detectability.
Detailed procedures are given in Appendix~\ref{app:methods}.

\begin{figure*}
\centering
\includegraphics[width=\textwidth]{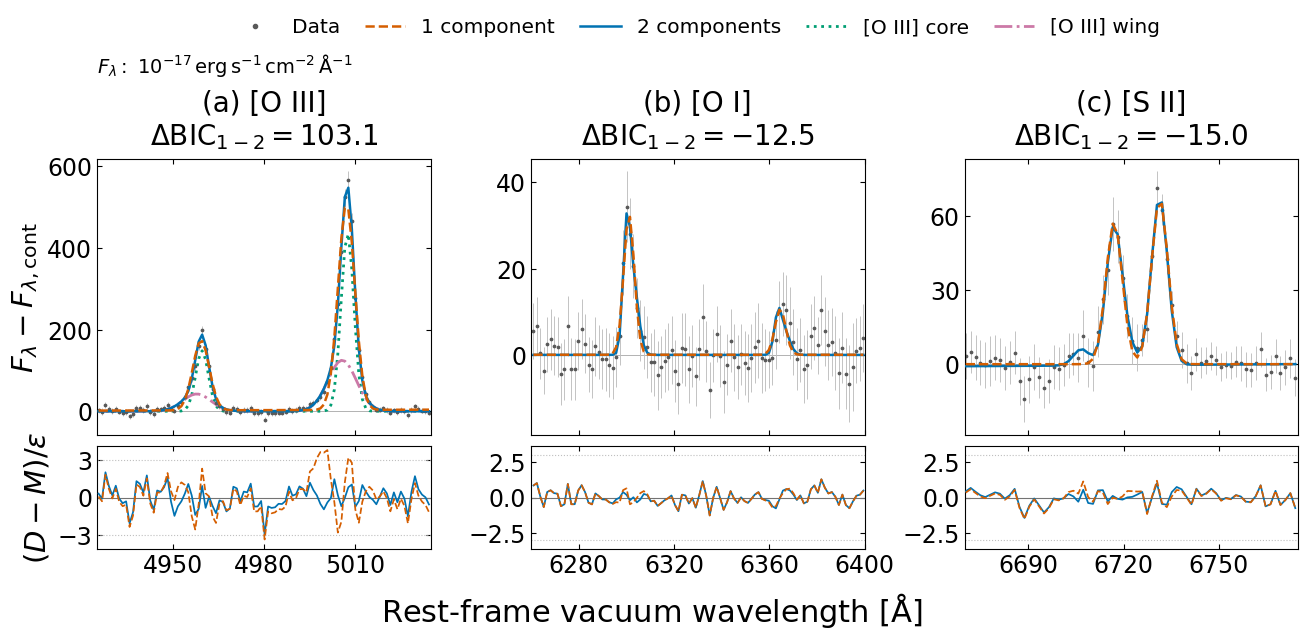}
\caption{
Independent fits to the 2024 forbidden-line profiles
for the degree-4 WDISP continuum.
Grey points and error bars show the data and nominal
uncertainties. Orange dashed and blue solid curves
show the one- and two-component models.
Green dotted and purple dash--dotted curves show
the fitted [O~III] core and wing.
The saved global continuum and the local background
of the adopted fit are subtracted from the data
and both models.
The adopted fits contain two components for [O~III]
and one for [O~I] and [S~II].
Lower panels show the normalised residuals.
Instrumental broadening and native-pixel integration
are included.
Positive $\Delta\mathrm{BIC}_{1-2}
=\mathrm{BIC}_{1}-\mathrm{BIC}_{2}$ favours the two-component fit.
}
\label{fig:ic1495_forbidden_profiles}
\end{figure*}

\subsection{Optical decomposition and narrow-line diagnostics}
\label{sec:nuclear_methods}
\label{sec:optical_methods}

The spectra are analysed independently on their native
grids, treating the 6dFGS arms separately where required
by their resolution and relative calibration.
E-MILES stellar templates \citep{Vazdekis2016} are fitted
with \textsc{ppxf} \citep{Cappellari2017}, testing nuclear
continua, optical Fe~II, polynomial orders, and
instrumental-resolution prescriptions.
Local emission-line fits include residual continua,
atomic doublet constraints, and instrumental broadening.
Appendices~\ref{app:6df_continuum_grid}
and~\ref{app:6df_line_models} describe the 6dFGS continuum
and line models, including masks, line-spread functions
(LSFs), and wavelength and velocity conventions.
The BOSS \texttt{BADFLUXFACTOR} OR-mask flag originates
from zero-weight blue-camera samples
(Appendix~\ref{app:sdssv_data}).

Excitation is assessed with BPT diagnostics
\citep{Baldwin1981} and the boundaries of
\citet{Kauffmann2003,Kewley2001,Kewley2006}.
WHAN \citep{CidFernandes2011} uses narrow
[N~II]~$\lambda6583$/H$\alpha$ and the rest-frame
narrow-H$\alpha$ equivalent width against the total
fitted local continuum.
The selected narrow-only solutions cover the tested
2004 configurations and the 2024 degree-4 and degree-8
WDISP continua.
The strong-AGN boundaries are
$\log_{10}([\mathrm{N\,II}]/\mathrm{H}\alpha)>-0.4$
and ${\rm EW}(\mathrm{H}\alpha)>6$~\AA.
These aperture-integrated diagnostics and their scenario
selection are detailed in
Appendix~\ref{app:boss_continuum_whan}.

For 6dFGS, we combine equivalent widths with SkyMapper
continuum estimates to reconstruct narrow-line fluxes
and the Balmer decrement. We vary the photometric
and aperture prescriptions and adopt an intrinsic
$({\rm H}\alpha/{\rm H}\beta)_0=3.10$ and the Fitzpatrick
extinction law with $R_V=3.1$ \citep{Fitzpatrick1999}.
The inferred attenuation applies to the narrow-line gas
under these assumptions. Absolute broad-line flux and
BLR and nuclear-continuum extinction remain unconstrained
by this reconstruction (Appendix~\ref{app:balmer_methods}).

The 2024 [O~III], [O~I], and [S~II] doublets are fitted
independently of H$\alpha$+[N~II], comparing one- and
two-component profiles across four continuum/LSF
configurations.
We compare the models using
$\Delta{\rm BIC}_{1-2}={\rm BIC}_1-{\rm BIC}_2$.
For the broader [O~III] component, or wing, we
measure its intrinsic FWHM, flux fraction, and
velocity relative to the core and independently fitted
[S~II].
The intrinsic mixture gives $W_{80}=v_{90}-v_{10}$
before instrumental convolution.
For the adopted two-component [O~III] and single-component
[S~II] models, 199 Gaussian parametric-bootstrap
realisations per configuration refit the line parameters
and local backgrounds.
The continuum, LSF, and component counts remain fixed.
We report conditional percentile intervals and the spread
across configurations.
The [S~II] ratio yields a single-zone electron density
with \textsc{PyNeb} \citep{Luridiana2015}, adopting
$T_{\rm e}=10^4$~K and testing alternative temperatures.
Out-of-grid ratios are retained as limits.
The narrow-line diagnostics are detailed in
Appendices~\ref{app:6df_line_models}
and~\ref{app:boss_narrow_kinematics}.

The supplementary [Ne~III] width test includes the
zero-width boundary. For [S~III], we compare free
principal kinematics with the same prescription plus
a fixed [O~III] wing. We leave the stellar velocity
dispersion undetermined because the accepted pixels
exclude the principal Mg~b and Fe absorption features
(Appendix~\ref{app:additional_line_profiles}).
Gas velocities use the reference frames of the optical fits.

\subsection{Broad Balmer emission and cross-epoch detectability}
\label{sec:broadline_methods}
\label{sec:hbeta_methods}
\label{sec:cross_epoch_methods}

We test broad H$\alpha$ using one shared H$\alpha$/[N~II]
component, independent narrow H$\alpha$ and [N~II]
kinematics, and two components with kinematics shared
between the lines. Each prescription is fitted with and
without a non-negative broad component.
All six models within a comparison use identical pixels
and weights, with 159 pixels in 2024 and 79 in 2004.
We refit line parameters and local residual backgrounds
at fixed global continuum.
Profiles include instrumental convolution and
native-pixel integration. The 2024 fitting window
and parameter bounds are given in
Appendix~\ref{app:sdssv_halpha_models}.

For each fixed configuration, the comparison statistic is
\begin{equation}
S_{\rm B}
=\min_{m\in\mathcal{M}_{\rm narrow}}{\rm BIC}_m
-\min_{m\in\mathcal{M}_{\rm narrow+broad}}{\rm BIC}_m,
\label{eq:sb_broad}
\end{equation}
where each minimum spans the three prescriptions and
${\rm BIC}_m=\chi_m^2+q_m\ln n$ for $q_m$ fitted
parameters and $n$ pixels.
Positive $S_{\rm B}$ favours the broad-containing family.
The reference calculation adopts the [N~II] energy-flux
ratio $R_{\rm NII}=F_{6583}/F_{6548}=3.05$,
consistent with the calculations and empirical mean
$3.049\pm0.021$ discussed by \citet{Dojcinovic2023NII}.
The alternative fixed ratio 2.96 tests sensitivity
to this constraint in the 2024 spectrum.

Extended 2024 tests combine four no-Fe~II continua
(WDISP/WRESL with multiplicative degrees 4/8),
local degrees 1/2, and $R_{\rm NII}=2.96$.
Additional prescriptions use [S~II]-fixed kinematics,
[S~II] plus a fixed [O~III] wing, or [S~II] plus
a free component, with independent H$\alpha$ and
[N~II] amplitudes.
The BIC differences compare paired fits with and
without broad H$\alpha$ within each prescription.
The null calibration applies to the reference six-model
comparison. To assess template sensitivity, we refit
the same H$\alpha$ data using 49 paired [S~II]/[O~III]
bootstrap template draws per continuum configuration.
We retain boundary solutions and zero broad amplitudes.

Conditional null simulations use each fitted narrow-only
model as a generator and repeat all six fits and model
selection. Gaussian simulations use the nominal pixel
errors. For empirical-noise simulations, we divide
control-window continuum residuals by their nominal
pixel errors, subtract the mean within each window,
and draw overlapping blocks of adjacent pixels with
replacement (moving-block resampling). We do not
rescale the centred residuals to unit variance.
This retains short-range structure within each block
and assumes that the control-window noise is
representative of the line window. Observed and
simulated spectra use matched multistart optimisation
and nested-model checks. Under correlated noise, we
use the BIC expression as a simulation-calibrated
comparison score.

Each generator has 199 realisations, with tail
probabilities $p=(k+1)/(N+1)$ and separately reported
Monte Carlo uncertainty.
The maximum $p_{\max}$ describes the conditional tails
of the three tested generators. Uniform false-positive
control over the composite null remains uncalibrated.
Appendix~\ref{app:sdssv_empirical_nulls} specifies
the 2024 configurations and the separate blue and
red noise controls.

Broad H$\beta$ is tested with the reference H$\alpha$
centroid and intrinsic width fixed, across four continua
and linear or quadratic local backgrounds.
Conditional null tests and injections compare Gaussian
and five-pixel moving-block noise.
We report the nominal broad-amplitude profile endpoint
at $\Delta\chi^2=2.71$. Its coverage and BLR extinction
remain undetermined.
The pixel selection, fixed parameters, and simulation
protocols are given in
Appendix~\ref{app:sdssv_hbeta_models}.

The 2004 recovery experiment uses the 2024 reference fit
with independent narrow H$\alpha$/[N~II]
kinematics, the degree-4 WDISP continuum, and
$R_{\rm NII}=3.05$. Its broad component has intrinsic
FWHM 1928.90~km~s$^{-1}$ and
$F_{\rm broad}/F_{\rm narrow}=0.8214$.
The injected component is scaled to each generator's
narrow-H$\alpha$ flux and centred on its narrow-line
velocity. Two-component generators use summed flux
and the flux-weighted centroid.
For each configuration, generator, noise prescription,
and injection level, 99 realisations preserve the earlier
sampling and LSF and repeat the full search with free
broad width and centroid.
The 2004 detection and recovery criterion is
$S_{\rm B}>0$ and $p_{\max}\leq0.05$
(equation~\ref{eq:recovery_criterion}).
Binomial intervals quantify recovery uncertainty,
and independent zero-injection trials assess
false-positive frequency.
Completeness is conditional on the simulated relative
strengths. Absolute broad-line flux transfer remains
undetermined under the tested unresolved-BLR and
extended-NLR aperture scenarios.
The 2004 fit bounds, null simulations, and injection
protocol are given in Appendix~\ref{app:injection_recovery}.

\subsection{X-ray spectral analysis}
\label{sec:nuclear_context_methods}
\label{sec:xray_spectral_methods}

The ungrouped \textit{Chandra}/ACIS spectrum is fitted
over 0.5--7.0~keV with Sherpa's W statistic, comparing
Galactic absorption alone with an additional absorber
at $z=0.02119$, the value used in the X-ray fits.
Models adopt the abundances of \citet{Wilms2000}
and cross-sections of \citet{Verner1996}.
We examine column-density--photon-index covariance
and fixed-$\Gamma$ models at 1.7, 1.8, and 1.9,
adopting $\Gamma=1.8$ as fiducial.
Column-density intervals use profile likelihoods.
Appendix~\ref{app:xray_methods} describes the fits,
luminosity estimates, and Fe~K$\alpha$ tests.

\subsection{Optical and MIR variability}
\label{sec:variability_analysis}

ZTF and WISE/NEOWISE analyses retain their AB and Vega
conventions (Appendix~\ref{app:variability_data}).
Colour--magnitude relations use covariance-aware
York regression \citep{York2004}.
Flux--flux relations use orthogonal-distance regression
to characterise correlated variable aperture emission.
Regression and optical--MIR lag-stability tests
are detailed in Appendix~\ref{app:chromatic_variability_methods}.
The IAC80 intraday sequences use the power-enhanced
$F$-test \citep{deDiego2014} at $\alpha=0.01$
(Appendix~\ref{app:iac80_variability_methods}).

\subsection{Host-galaxy structure and atomic gas}
\label{sec:host_galaxy_methods}
\label{sec:host_methods}
\label{sec:hi_methods}

DESI Legacy Imaging Surveys DR10 $g,r,z$ images
\citep{Dey2019} provide integrated photometry,
outer-disc geometry, and deprojected stellar structure.
Isophotal fits define the reference geometry, and
stellar masses use colour-dependent mass-to-light
calibrations (Appendix~\ref{app:host_photometry_methods}).
Fourier analysis measures the bar and outer asymmetry
(Appendix~\ref{app:fourier_tests}).
PSF-convolved two-dimensional models provide a
complementary structural check
(Appendix~\ref{app:imfit_methods}).

The Nançay H\,I profile \citep{Courtois2009}
provides integrated flux, linewidth, and velocity-space
asymmetry.
Correlated-noise simulations and alternative windows
and baselines assess uncertainty.
Symmetric-profile simulations calibrate the
asymmetry statistic.
Integrated flux yields the atomic-gas mass,
while the EDD linewidth and optical inclination
provide a rotation-velocity proxy.
Stellar and H\,I mass calculations retain their
shared distance dependence.
Appendix~\ref{app:hi_methods} gives the definitions
and the stellar-plus-atomic-gas baryonic Tully--Fisher
comparison.

\section{Results}
\label{sec:results}

\subsection{Nuclear emission, gas kinematics, and broad-line constraints}
\label{sec:ic1495_nuclear_results}

\subsubsection{Host dilution and Seyfert excitation}
\label{sec:ic1495_optical_nuclear_results}

Host+AGN models are preferred throughout the 2004 continuum grid.
Stellar light dominates near 5100~\AA\ and H$\alpha$
(Table~\ref{tab:ic1495_optical_summary}), and the adopted fits
omit optical Fe~II. The continuum and photometric sensitivity
tests are summarised in Table~\ref{tab:ic1495_optical_robustness},
and the reference decomposition is shown in
Fig.~\ref{fig:ic1495_continuum}
(Appendix~\ref{app:optical_results_robustness}).

The 2004 BPT ratios place IC~1495 in the AGN region
of the [N~II] plane and on the Seyfert branches of the
[S~II] and [O~I] planes
(Table~\ref{tab:ic1495_optical_summary}).
The [O~I]~$\lambda6300$ detection gives
$\Delta{\rm BIC}=28.9$ and
$p_{\rm MC}=1/3001\simeq3.3\times10^{-4}$
(Appendix~\ref{app:6df_line_models}).
All accepted [O~I] profiles place the source on the Seyfert
branch (Table~\ref{tab:ic1495_optical_robustness}).
The selected narrow-only fits place both epochs in the
strong-AGN region of the WHAN diagram
(Appendix~\ref{app:boss_continuum_whan}).
Aperture and continuum differences limit the interpretation
of the displacement between epochs.

The reconstructed narrow-line decrement,
$({\rm H}\alpha/{\rm H}\beta)_{\rm phot}=7.10^{+0.90}_{-0.74}$,
implies substantial NLR reddening under the adopted extinction law
and intrinsic decrement (Table~\ref{tab:ic1495_optical_summary}).

\begin{table}
\centering
\caption{Reference continuum and narrow-line measurements from the 2004 6dFGS spectrum.}
\label{tab:ic1495_optical_summary}
\footnotesize
\setlength{\tabcolsep}{5pt}
\renewcommand{\arraystretch}{1.15}
\begin{tabular}{ll}
\hline
Quantity & Value \\
\hline
$f_{\rm host}(5100)$ & 0.671--0.704 \\
$f_{\rm host}({\rm H}\alpha)$ & 0.735--0.815 \\
$\log([\mathrm{O\,III}]/\mathrm{H}\beta)$ &
$0.688^{+0.045}_{-0.040}$ \\
$\log([\mathrm{N\,II}]/\mathrm{H}\alpha)$ &
$-0.002^{+0.016}_{-0.017}$ \\
$\log([\mathrm{S\,II}]/\mathrm{H}\alpha)$ &
$-0.360^{+0.034}_{-0.035}$ \\
$\log([\mathrm{O\,I}]/\mathrm{H}\alpha)$ &
$-1.015^{+0.100}_{-0.116}$ \\
$E(B-V)_{\rm NLR}$ &
$0.716^{+0.103}_{-0.094}$ mag \\
$n_{\rm e}$ at $10^4$ K &
$261^{+196}_{-157}$ cm$^{-3}$ \\
$\log L_{[\mathrm{O\,III}]}$ (erg s$^{-1}$) &
40.184 \\
\hline
\end{tabular}
\par\smallskip
\begin{minipage}{\columnwidth}
\footnotesize\raggedright
\textit{Notes.} The stellar fractions span the tested
continuum prescriptions. The line-ratio intervals give the
16th--84th percentiles from Monte Carlo refitting
(Appendix~\ref{app:6df_line_models}).
The ratios use [O~III]~$\lambda5007$, [N~II]~$\lambda6583$,
the summed [S~II] doublet, and [O~I]~$\lambda6300$.
The [O~III] luminosity includes Galactic foreground
correction only. The WHAN analysis uses ratios from a
separate set of selected narrow-only fits
(Appendix~\ref{app:boss_continuum_whan}).
\end{minipage}
\end{table}

\subsubsection{Independent narrow-line kinematics and gas density}
\label{sec:ic1495_narrow_kinematics_results}

Across four continuum/LSF configurations, the 2024 [O~III] profile
favours two components ($\Delta{\rm BIC}_{1-2}=103.13$--103.20).
Single-component fits are preferred for [O~I]
($-12.57$ to $-12.54$) and [S~II] (approximately $-14.98$)
at the achieved sensitivity. The fits and residuals are shown in
Fig.~\ref{fig:ic1495_forbidden_profiles}.

The [O~III] wing contributes approximately 39 per cent of the line
flux and is broader and blueshifted relative to the core and [S~II]
(Table~\ref{tab:ic1495_narrow_2024}). The intrinsic total-profile
$W_{80}$ spans 439.7--446.0~km~s$^{-1}$ across configurations.
All 199 bootstrap trials per configuration yield valid fits without
recorded kinematic-boundary solutions. The intervals are conditional
on the adopted continuum, LSF, component count, and Gaussian noise
(Appendix~\ref{app:boss_narrow_kinematics}).

At $T_{\rm e}=10^4$~K, the [S~II] density has a 2.5th--97.5th
percentile range of 590--2918~cm$^{-3}$. The point estimates at
8000, 15\,000, and 20\,000~K are 1137, 1452, and 1623~cm$^{-3}$,
respectively. These estimates apply to the [S~II]-emitting gas.
Appendix~\ref{app:additional_line_profiles} reports the [Ne~III] and
[S~III] tests, the search near 6382~\AA, and the stellar fits.
The latter leave the stellar velocity dispersion undetermined.

\begin{table}
\centering
\caption{Independent narrow-line measurements from the 2024 SDSS-V/BOSS spectrum.}
\label{tab:ic1495_narrow_2024}
\footnotesize
\setlength{\tabcolsep}{3pt}
\renewcommand{\arraystretch}{1.15}
\begin{tabular}{lrr}
\hline
Quantity & Best fit & Percentile range \\
\hline
$F_{6716}/F_{6731}$ &
0.859 & 0.750--0.939 \\
${\rm FWHM}_{[\mathrm{O\,III}],{\rm wing}}$ &
638 & 596--674 \\
$v_{\rm wing}-v_{\rm core}$ &
$-105$ & $-129$ to $-88$ \\
$v_{\rm wing}-v_{[\mathrm{S\,II}]}$ &
$-91$ & $-118$ to $-73$ \\
$F_{\rm wing}/F_{[\mathrm{O\,III}],{\rm total}}$ &
0.388 & 0.343--0.443 \\
$W_{80,[\mathrm{O\,III}]}$ &
446 & 437--459 \\
$n_{\rm e}$ (cm$^{-3}$) &
1245 & 906--1983 \\
\hline
\end{tabular}
\par\smallskip
\begin{minipage}{\columnwidth}
\footnotesize\raggedright
\textit{Notes.} Values use the degree-4 WDISP reference
configuration. Intervals are the conditional 16th--84th
percentiles from 199 Gaussian bootstrap trials.
Velocities and intrinsic widths are in km~s$^{-1}$.
The density assumes $T_{\rm e}=10^4$~K, and its interval
excludes atomic-data systematics.
\end{minipage}
\end{table}

\subsubsection{Broad Balmer constraints and cross-epoch detectability}
\label{sec:ic1495_broadline_2004_2024}
\label{sec:ic1495_hbeta_results}

None of the eight 2004 configurations satisfies the adopted
broad-H$\alpha$ detection criterion (equation~\ref{eq:recovery_criterion}).
Table~\ref{tab:ic1495_broadline_summary} summarises the comparisons
and Table~\ref{tab:ic1495_broadline_robustness} gives the full 2004 grid.
The 2024 spectrum admits a broad-H$\alpha$-like excess whose support
depends on the decomposition. The same-exposure COADD and
\texttt{MJD\_EXP} products give consistent reference scores
(Appendix~\ref{app:broadline_results_robustness}).
Across matching continuum/LSF configurations, changing $R_{\rm NII}$
from 3.05 to 2.96 lowers $S_{\rm B}$ by approximately one.
Changes from the continuum degree and crossed emission-line LSF
are 0.10--0.11 and 0.002--0.003, respectively.

In the empirical block runs, null-noise realizations
are generated by resampling contiguous groups of
error-normalized continuum residuals from the blue
or red control windows
(Appendix~\ref{app:sdssv_empirical_nulls}).
At $R_{\rm NII}=3.05$, the reference Gaussian and blue-control
block runs give $p_{\max}=0.005$, the finite-ensemble floor.
The five-pixel red-control run gives 0.020. At $R_{\rm NII}=2.96$,
the two-component narrow-only generator gives 11/199 and 13/199
Gaussian exceedances for the degree-4 and degree-8 WDISP continua,
respectively. The corresponding 95-per-cent binomial intervals are
0.028--0.097 and 0.035--0.109. Strengthened multistart fits reproduce
these counts. The blue- and red-control block probabilities lie on
opposite sides of 0.05 (Table~\ref{tab:ic1495_broadline_summary}).
The blue and red control residuals differ in measured
RMS amplitude and autocorrelation pattern. The
probabilities are conditional on the noise prescription
and fitted generators.

\begin{figure*}
\centering
\includegraphics[width=0.90\textwidth]{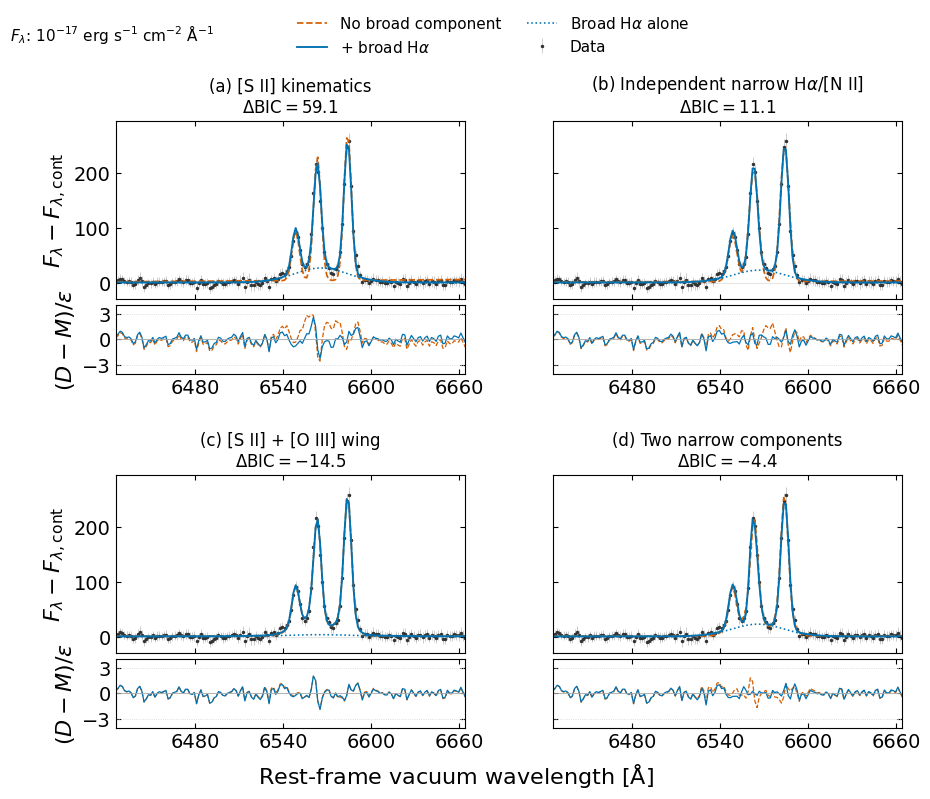}
\caption{Alternative descriptions of the 2024 H$\alpha$+[N~II]
complex. Panel (a) uses fixed [S~II] kinematics, (b) independent
narrow H$\alpha$/[N~II] kinematics, (c) [S~II] plus a fixed
[O~III] wing, and (d) two free components with kinematics
shared between H$\alpha$ and [N~II].
Points show continuum-subtracted data with nominal uncertainties.
Orange dashed and blue solid curves show fits without and with
broad H$\alpha$, respectively, and blue dotted curves show the
broad component. Lower strips show normalised residuals.
All panels use the degree-4 WDISP continuum, a linear local
background, and $R_{\rm NII}=2.96$.
Positive $\Delta{\rm BIC}={\rm BIC}_{\rm no\ broad}
-{\rm BIC}_{+\rm broad}$ favours adding broad H$\alpha$
within the prescription shown in that panel.
External-template sensitivity is evaluated separately.
Flux densities retain the observed-frame convention after
Galactic-extinction correction, and wavelengths are rest-frame
vacuum.}
\label{fig:ic1495_halpha_models}
\end{figure*}

The largest changes in broad-line support arise from the narrow-line
kinematics (Fig.~\ref{fig:ic1495_halpha_models}). Across four continua
and two local background degrees at $R_{\rm NII}=2.96$, the paired
$\Delta{\rm BIC}_{\rm narrow-broad}$ spans 49.74--60.75 with fixed
[S~II] kinematics and $-14.55$ to $-12.77$ when a fixed [O~III] wing is added.
With fixed [S~II] and a free second component it spans
$-12.59$ to $-11.67$, with some broad solutions at parameter bounds.
For freely fitted kinematics, the ranges are 6.85--7.90 for one
shared component, 11.11--11.73 for independent narrow
H$\alpha$/[N~II], and $-4.53$ to $-2.89$ for two shared components.
Table~\ref{tab:ic1495_ha_narrow_components} gives the narrow-only
parameters of the models shown in the figure.

Among the six free-kinematic fits for the configuration of
Fig.~\ref{fig:ic1495_halpha_models}, the lowest BIC is obtained with
broad H$\alpha$ and independent narrow H$\alpha$/[N~II] kinematics.
The best narrow-only fit has two shared components. Comparing
these minima gives $S_{\rm B}=4.05$
(Table~\ref{tab:ic1495_ha_free_model_bic}).
Within the two-shared-component prescription, adding broad
H$\alpha$ raises the BIC by 4.43.

In each continuum configuration, all 49 template perturbations favour
adding broad H$\alpha$ for the fixed [S~II] prescription and favour
omitting it when the fixed [O~III] wing is included. The [S~II]-only
broad solutions have positive amplitudes and interior parameters.
With the fixed wing, 18--19 trials per configuration reach parameter
bounds and 2--3 give zero broad amplitude. These tests vary the
templates at fixed H$\alpha$ data
(Appendix~\ref{app:sdssv_halpha_models}).

With independent narrow H$\alpha$/[N~II] kinematics, a quadratic
local background increases the broad FWHM from approximately
1930 to 2110~km~s$^{-1}$. The extended grid gives
$S_{\rm B}=3.95$--5.52 for the free prescriptions. Noise calibration
applies to the configurations in Appendix~\ref{app:sdssv_empirical_nulls}.
The calibration audit traces the \texttt{BADFLUXFACTOR} OR bit to
zero-weight blue-camera samples (Appendix~\ref{app:sdssv_data}).
The sensitivity to narrow-line kinematics and residual noise leaves
the excess as a BLR candidate.

\begin{figure*}
\centering
\includegraphics[width=\textwidth]{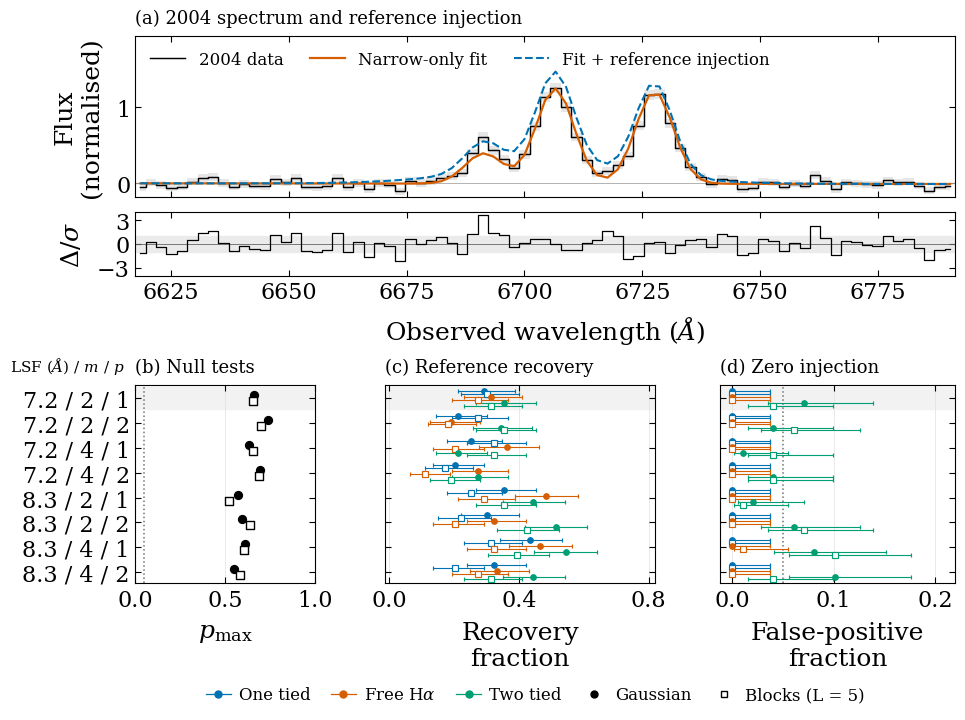}
\caption{Broad-H$\alpha$ detectability in 2004.
Panel (a) shows the continuum-subtracted data, the narrow-only
fit, and the fit plus the reference injection with intrinsic
FWHM $\simeq1929$~km~s$^{-1}$ and
$F_{\rm broad}/F_{\rm narrow}=0.8214$.
The grey band shows nominal uncertainties, and the lower
strip gives normalised narrow-only residuals.
Panels (b)--(d) show the maximum conditional null probabilities,
reference recovery fractions, and zero-injection false-positive
fractions, respectively.
Rows give the observed-frame LSF FWHM in \AA\ and polynomial
degrees $m,p$. The shaded row gives the configuration in
panel (a).
Colours identify the narrow-line generators, and filled circles
and open squares indicate Gaussian and five-pixel block noise,
respectively.
Fraction intervals are 95-per-cent Wilson intervals for 99
trials per cell. The dotted lines in panels (b) and (d) mark 0.05.}
\label{fig:ic1495_broadline_2004_2024}
\end{figure*}

For an intrinsic broad-line FWHM of 1928.90~km~s$^{-1}$,
baseline recovery is 3--18, 27--35, and 42--55 per cent at
injected broad-to-narrow H$\alpha$ flux ratios of 0.411, 0.821,
and 1.232, respectively (Table~\ref{tab:ic1495_broadline_summary}).
The ranges span three narrow-line generators and two noise
prescriptions. At the reference ratio of 0.821, recovery spans
11--55 per cent across the eight configurations
(Fig.~\ref{fig:ic1495_broadline_2004_2024}).
Two zero-injection cells give 10/99 detections each, with a
95-per-cent Wilson interval of 5.6--17.6 per cent.
These rates apply to the tested generators and noise prescriptions.

Even a component at the largest tested baseline ratio could have
escaped detection in 2004. The aperture scenarios in
Appendix~\ref{app:broadline_results_robustness} show how that ratio
depends on the assumed narrow-line extent. Aperture and flux-calibration
uncertainties leave the absolute broad-line flux change undetermined.
The two spectra do not establish a changing-look transition.

The fixed-profile tests favour omitting broad H$\beta$, with
$\Delta{\rm BIC}_{\rm narrow-broad}$ from $-4.95$ to $-4.20$ and
reference null probabilities of 0.330--0.425. The nominal conditional
Balmer-ratio endpoints are 5.13--5.93, with up to 16 per cent
variation in the H$\beta$ amplitude endpoint across continua.
For injected broad H$\alpha$/H$\beta$ ratios of 3.1, 5, and 10,
recovery is 91--100, 66--83, and 28--38 per cent, respectively.
The endpoints condition on the adopted H$\alpha$ profile and have
uncalibrated coverage. The BLR extinction remains undetermined
(Appendix~\ref{app:sdssv_hbeta_models}).

\begin{table}
\centering
\caption{Broad-H$\alpha$ constraints and injection--recovery results for IC~1495.}
\label{tab:ic1495_broadline_summary}
\footnotesize
\setlength{\tabcolsep}{3pt}
\renewcommand{\arraystretch}{1.15}
\begin{tabular}{lll}
\hline
Test & Quantity & Result \\
\hline
2004 model grid & $S_{\rm B}$ & $-2.92$ to $+1.65$ \\
2004 null tests & $p_{\max}$ & 0.52--0.74 \\
2024, $R_{\rm NII}=3.05$ & $S_{\rm B}$ & 4.99--5.10 \\
2024, $R_{\rm NII}=2.96$ & $S_{\rm B}$ & 3.95--4.06 \\
Gaussian, $m=4$ & $p_{\max}$ & 0.060 \\
Gaussian, $m=8$ & $p_{\max}$ & 0.070 \\
Blue blocks, $m=4$ & $p_{\max}$ & 0.045 \\
Blue blocks, $m=8$ & $p_{\max}$ & 0.020 \\
Red blocks, $m=4$ & $p_{\max}$ & 0.110 \\
Red blocks, $m=8$ & $p_{\max}$ & 0.140 \\
2024 linear background & FWHM & 1923--1929 km~s$^{-1}$ \\
2024 linear background & $r$ & 0.803--0.829 \\
2024 quadratic background$^{a}$ & FWHM & 2103--2117 km~s$^{-1}$ \\
Reference injection & FWHM & $\simeq1929$ km~s$^{-1}$ \\
Reference injection & $r$ & 0.8214 \\
Recovery, full grid & $P_{\rm rec}(0.821)$ & 0.11--0.55 \\
Recovery, baseline & $P_{\rm rec}(0.411)$ & 0.03--0.18 \\
                    & $P_{\rm rec}(0.821)$ & 0.27--0.35 \\
                    & $P_{\rm rec}(1.232)$ & 0.42--0.55 \\
\hline
\end{tabular}
\par\smallskip
\begin{minipage}{\columnwidth}
\footnotesize\raggedright
\textit{Notes.} Positive $S_{\rm B}$ favours the best
broad-containing fit over the best narrow-only fit.
The 2024 linear-background fits use matching continuum
and emission-line LSFs. The 2024 null tests listed here
use $R_{\rm NII}=2.96$ and WDISP. The quantity $p_{\max}$
is the maximum conditional null probability over three
fitted narrow-only generators. The continuum degree is $m$,
and moving blocks contain five pixels.
The flux ratio is $r=F_{\rm broad}/F_{\rm narrow}$.
The recovery fraction $P_{\rm rec}(r)$ is the fraction of
injection trials that meet equation~\ref{eq:recovery_criterion}.
The recovery ranges span the generators and noise
prescriptions. The full grid also varies the continuum,
local background, and LSF.
$^{a}$Independent narrow H$\alpha$/[N~II] fits at $R_{\rm NII}=2.96$.
\end{minipage}
\end{table}

\subsubsection{X-ray obscuration}
\label{sec:ic1495_xray_results}

The point-like \textit{Chandra} source coincides with the optical
nucleus and has approximately 74 net counts over 0.5--7.0~keV,
nearly all above 2~keV ($\mathrm{HR}\simeq0.95$).
The Galactic-only fit gives $\Gamma\simeq-1.62$.
Adding intrinsic absorption improves the free-$\Gamma$ fit by
$\Delta W\simeq8.1$. The strong covariance between $N_{\rm H,int}$
and $\Gamma$ is shown in Fig.~\ref{fig:ic1495_nh_gamma}
(Appendix~\ref{app:xray_results_robustness}).

At fixed $\Gamma=1.8$, the fiducial model gives
$N_{\rm H,int}=2.25^{+0.70}_{-0.57}\times10^{23}$~cm$^{-2}$
(90-per-cent profile-likelihood interval) and
$\log_{10}[L_{2-10}^{\rm int}/(\mathrm{erg\,s^{-1}})]=42.45$.
The luminosity extrapolates beyond the observed 7-keV fitting limit.
The tested fixed slopes give comparable columns in the heavily
absorbed, Compton-thin regime (Table~\ref{tab:ic1495_xray_robustness}).
The narrow Fe~K$\alpha$ search gives equivalent-width upper limits
in the same appendix.

\subsubsection{Optical and MIR variability}
\label{sec:ic1495_variability_results}

\begin{figure}
\centering
\includegraphics[width=0.90\columnwidth]{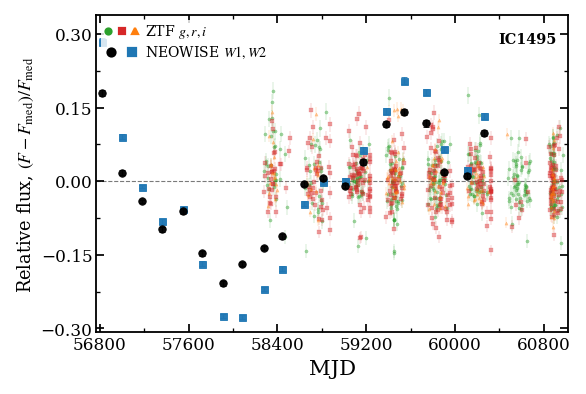}
\caption{
ZTF and WISE/NEOWISE light curves of IC~1495.
The observed MIR magnitude amplitude exceeds
that of the optical light curves.
}
\label{fig:ic1495_longterm}
\end{figure}

\begin{figure*}
\centering
\begin{minipage}{0.49\textwidth}
\centering
\includegraphics[width=0.90\linewidth]{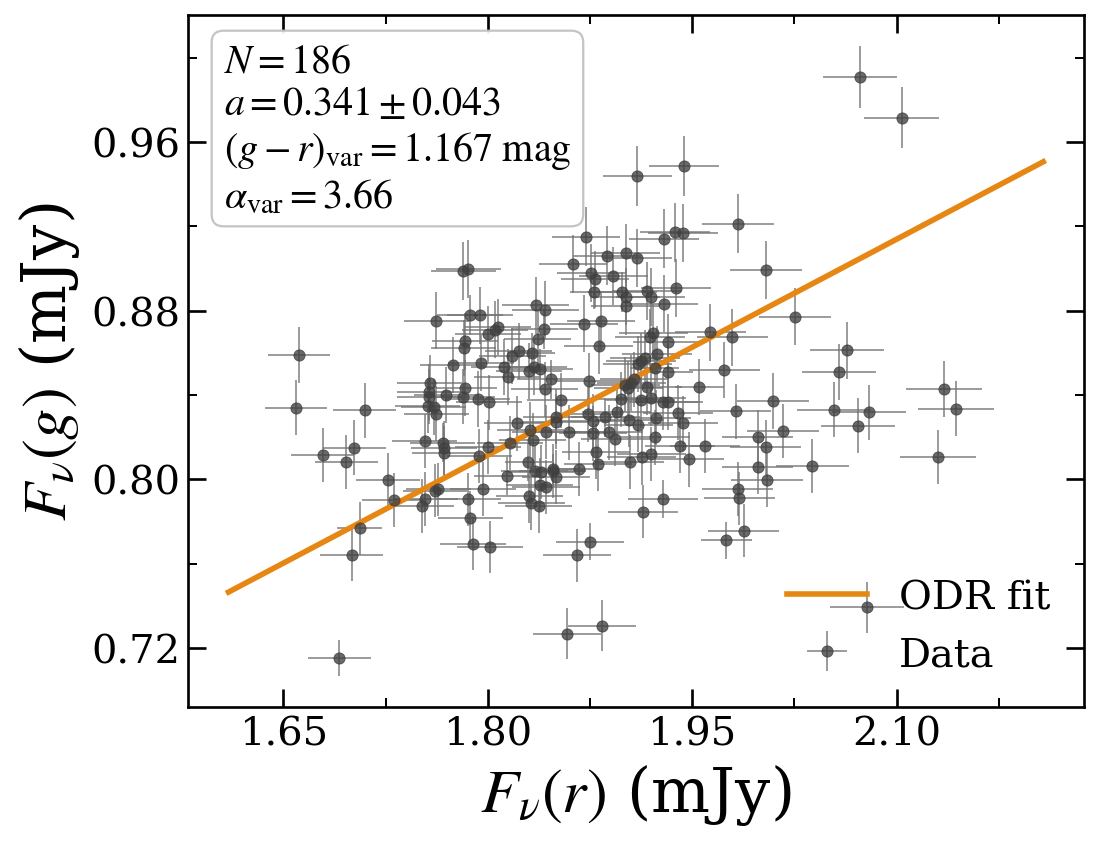}
\end{minipage}
\hfill
\begin{minipage}{0.49\textwidth}
\centering
\includegraphics[width=0.90\linewidth]{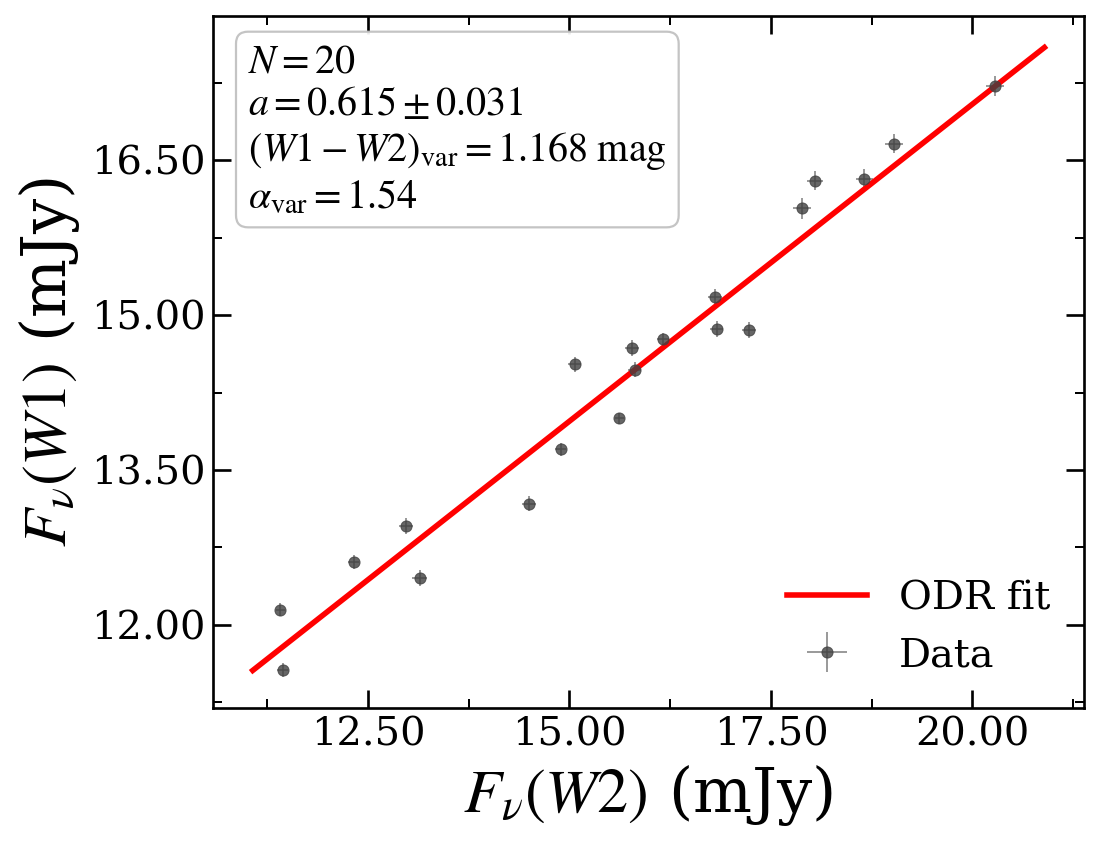}
\end{minipage}
\caption{
Flux--flux relations for ZTF $g,r$ measurements
(left) and WISE/NEOWISE $W1,W2$ epoch averages
(right). Slopes describe the effective colours
of the correlated variable aperture emission.
}
\label{fig:ic1495_chromatic}
\end{figure*}

The aperture-integrated light curves show larger observed magnitude
variations in the MIR than in the optical
(Fig.~\ref{fig:ic1495_longterm}). Colour--magnitude fits show
optical bluer-when-brighter and MIR redder-when-brighter trends.
Table~\ref{tab:ic1495_chromatic_full} gives the slopes and effective
variable colours for all band pairs. The principal flux--flux
relations are shown in Fig.~\ref{fig:ic1495_chromatic}.

The optical--MIR correlation is broad and the resampled centroid
distribution is multimodal, leaving the lag and dust radius
undetermined (Appendix~\ref{app:chromatic_variability_methods}).
The IAC80 sequences fall below the adopted intraday-variability
threshold (Table~\ref{tab:ic1495_idv_results},
Appendix~\ref{app:ic1495_idv}).

\subsection{Host-galaxy structure and atomic gas}
\label{sec:ic1495_host_results}
\label{sec:ic1495_host_geometry_results}
\label{sec:ic1495_fourier_results}
\label{sec:ic1495_hi_results}

The outer-disc geometry stabilises beyond approximately
$40^{\prime\prime}$ and defines the deprojection and H\,I
inclination correction (Table~\ref{tab:ic1495_host_summary}).
The fiducial stellar mass is
$M_\star=6.41^{+2.48}_{-1.80}\times10^{10}\,M_\odot$.
Table~\ref{tab:ic1495_host_robustness} gives the supplementary
photometric and structural tests. The stellar-mass prescriptions are
given in Appendix~\ref{app:host_photometry_methods}.

\begin{figure}
\centering
\includegraphics[width=0.90\columnwidth]{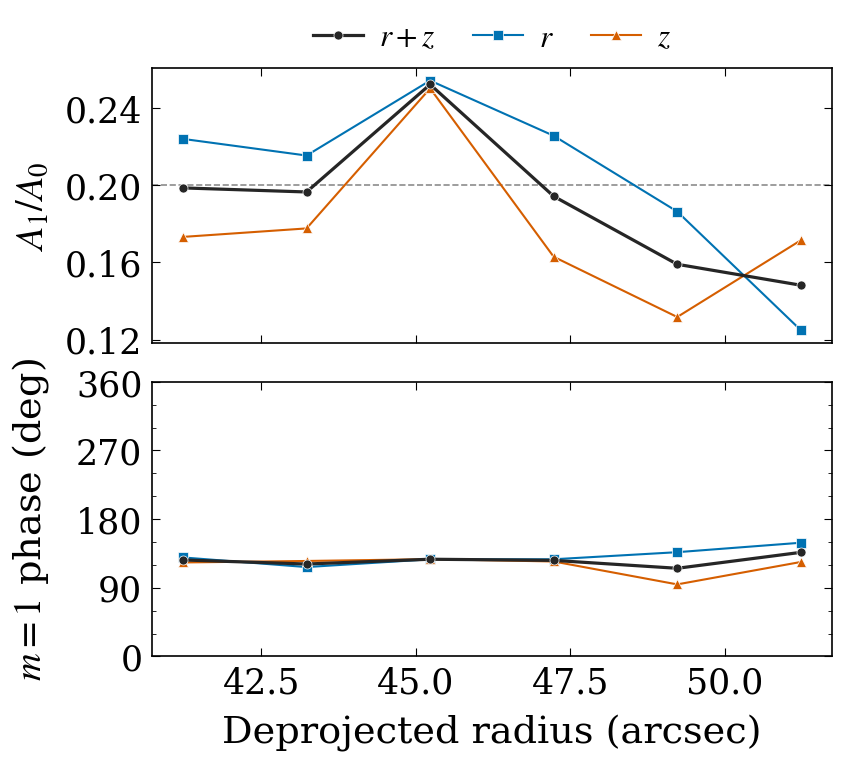}
\caption{
Outer stellar $m=1$ structure.
Upper and lower panels show the amplitude
and phase in $r$, $z$, and their combined
measurement. The amplitude is approximately
0.20 with coherent phase across the adopted
radial interval.
}
\label{fig:ic1495_m1}
\end{figure}

A phase-coherent $m=2$ mode defines the deprojected bar-radius
bracket of 9.65--11.26~kpc (Fig.~\ref{fig:ic1495_bar}).
At $40$--$51$~arcsec, the combined $r+z$ profile has a coherent
outer $m=1$ mode with $A_1/A_0\simeq0.20$
(Fig.~\ref{fig:ic1495_m1}). None of the 5000 azimuth-scrambled
trials reaches the observed amplitude--coherence statistic, giving
$p_{\rm MC}=1/5001\simeq2.0\times10^{-4}$ under that null model
(equation~\ref{eq:finite_mc_probability}). The mode persists
under sector omission and image reconstruction
(Appendix~\ref{app:host_results_robustness}).

\begin{figure}
\centering
\includegraphics[width=\columnwidth]{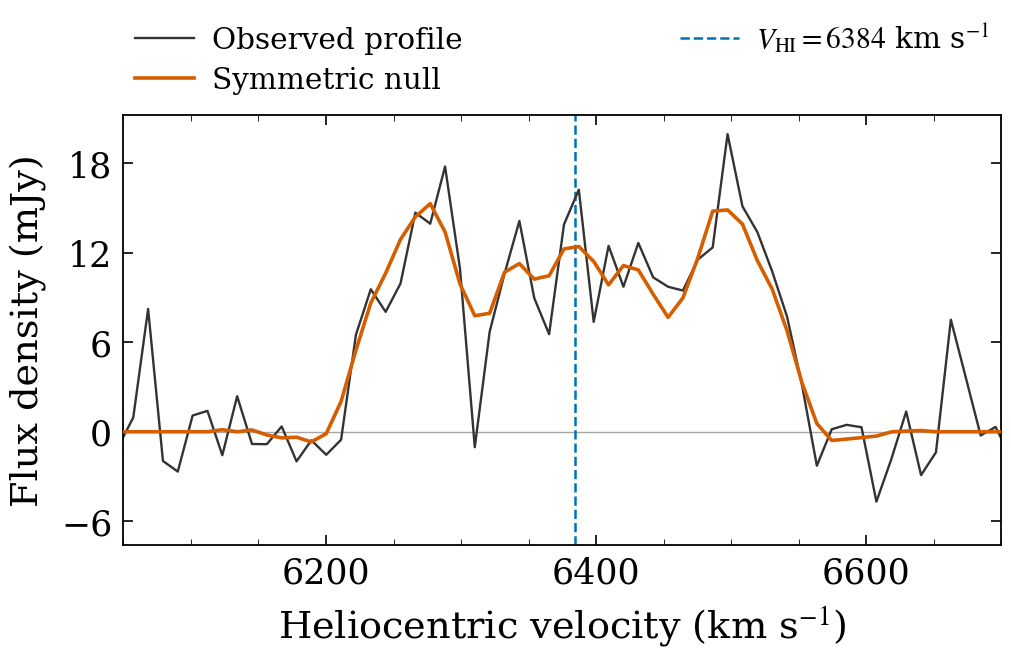}
\caption{
Global Nançay H\,I profile, symmetric null
template, and adopted systemic velocity.
The symmetry test gives $p_{\rm sym}=0.186$.
}
\label{fig:ic1495_hi_profile}
\end{figure}

The remeasured Nançay H\,I profile agrees with the catalogue
measurements. Table~\ref{tab:ic1495_host_summary} gives the integrated
flux, inferred H\,I mass, and gas-to-stellar mass ratio. We adopt the
EDD linewidth $W_{mx}=304\pm18$~km~s$^{-1}$ for linewidth-based
quantities. The $|A_{\rm signed}|$ statistic
(equation~\ref{eq:hi_asymmetry}) gives $p_{\rm sym}=0.186$ under the
symmetric-profile null (Fig.~\ref{fig:ic1495_hi_profile}).
Table~\ref{tab:ic1495_hi_robustness} gives the linewidth and
baryonic Tully--Fisher checks. Beam screening finds no substantial
known velocity-matched contaminant. The redshift of one unresolved
in-beam source remains poorly constrained
(Appendix~\ref{app:hi_confusion}).

\begin{table}
\centering
\caption{Principal host-galaxy and global H\,I measurements for IC~1495.}
\label{tab:ic1495_host_summary}
\footnotesize
\setlength{\tabcolsep}{5pt}
\renewcommand{\arraystretch}{1.15}
\begin{tabular}{ll}
\hline
Quantity & Value \\
\hline
$q_{\rm out}$ & $0.705\pm0.016$ \\
$i$ & $46.3\pm1.4^\circ$ \\
$\log(M_\star/M_\odot)$ &
$10.807^{+0.142}_{-0.143}$ \\
$(A_2/A_0)_{\rm max}$ &
$0.804^{+0.015}_{-0.017}$ \\
$R_{\rm bar}$ & 9.65--11.26 kpc \\
$(A_1/A_0)_{\rm outer}$ &
$0.199^{+0.014}_{-0.010}$ \\
$p_{\rm MC}(m=1)$ &
$\simeq2.0\times10^{-4}$ \\
$F_{\rm HI}$ &
$3.60\pm0.29$ Jy km s$^{-1}$ \\
$M_{\rm HI}$ &
$7.03^{+1.31}_{-1.17}\times10^9\,M_\odot$ \\
$M_{\rm HI}/M_\star$ &
$0.109^{+0.045}_{-0.033}$ \\
$A_{\rm signed}$ &
$0.079^{+0.087}_{-0.068}$ \\
$p_{\rm sym}$ & 0.186 \\
\hline
\end{tabular}
\par\smallskip
\begin{minipage}{\columnwidth}
\footnotesize\raggedright
\textit{Notes.} The bar-radius interval is the deprojected
Fourier bracket. The $m=1$ probability is conditional on
the azimuth-scrambling null.
\end{minipage}
\end{table}

\section{Discussion}
\label{sec:discussion}

\subsection{An absorbed variable nucleus with weak optical signatures}
\label{subsec:discussion_nuclear_context}

Seyfert narrow-line excitation, compact X-ray emission and long-term
variability establish an active nucleus independently of the
broad-H$\alpha$ candidate. The X-ray column and the narrow-line
reddening probe different regions, sightlines and epochs, leaving
BLR attenuation and changes in obscuration undetermined.

Stellar dilution limits broad-line contrast in the 2004 spectrum.
The detection of broad H$\alpha$ in NGC~3147 with a narrow HST
aperture illustrates the importance of host contrast
\citep{Bianchi2019NGC3147}. The stellar fractions measured in IC~1495
apply to the larger 6dFGS aperture.

A constant host and a variable component of fixed colour can
produce the observed aperture-integrated colour trends. The
flux--flux slopes characterise the colour of the correlated
variability. The MIR changes are consistent with variable dust
reprocessing, and the multimodal lag distribution leaves the
reverberation radius undetermined.

The broad-H$\beta$ tests constrain detectability for the adopted
profile (Section~\ref{sec:ic1495_hbeta_results}). If the H$\alpha$
excess includes BLR emission, attenuation, intrinsically weak
emission and differential Balmer-line responsivity remain possible
explanations for the weak H$\beta$ signature
\citep{Trippe2010,Gaskell2023Reddening,Son2025Balmer,Guo2025b}.

\subsection{Ionized-gas kinematics and template transfer}
\label{subsec:discussion_gas_kinematics}

The broader, blueshifted [O~III] component is recovered across the
tested continuum and LSF configurations. The preference for simpler
[S~II] and [O~I] profiles applies at the sensitivity of each line
(Fig.~\ref{fig:ic1495_forbidden_profiles}). Wing strength and width
can vary between ions \citep{KovacevicDojcinovic2022}. Transferring
the [O~III] wing to H$\alpha$ and [N~II] therefore assumes shared
kinematics, which the ancillary line tests leave unverified
(Appendix~\ref{app:additional_line_profiles}).

An outflow could produce the [O~III] asymmetry. The unresolved
spectrum and limited stellar-kinematic constraints leave the
contribution of gravitational motions uncertain. The intrinsic
$W_{80}$ describes the line profile. Estimating the mass and
energetics of a putative outflow would require the wing density,
extent and geometry. The [S~II] ratio constrains the density of
its own emitting gas \citep{Davies2020Densities}.

\subsection{Broad-line ambiguity and limits on spectral evolution}
\label{subsec:discussion_nuclear_state}

Among the six free-kinematic fits in the configuration of
Fig.~\ref{fig:ic1495_halpha_models}, the lowest BIC is obtained
with broad H$\alpha$ and independent narrow H$\alpha$/[N~II]
kinematics (Table~\ref{tab:ic1495_ha_free_model_bic}).
Within the prescription with two shared narrow components,
adding broad H$\alpha$ increases the BIC. The narrow-only fits
show how additional velocity structure can account for the
excess (Table~\ref{tab:ic1495_ha_narrow_components}).
This dependence persists under perturbations of the external
templates. Similar effects of narrow-line modelling occur in
nearby low-luminosity AGNs and X-ray-selected intermediate
Seyferts \citep{BalmaverdeCapetti2014,Shimizu2018BASSVIII}.
The model and noise sensitivity leaves the broad H$\alpha$
component a BLR candidate and the Seyfert~1.9 classification
unconfirmed.

The recovery fractions in Fig.~\ref{fig:ic1495_broadline_2004_2024}
show that the reference profile could have escaped detection in
2004. They apply to the specified broad-to-narrow flux ratio and
profile. Aperture transfer and flux-calibration uncertainties
leave the absolute change in broad-line flux undetermined.

The classifications of \citet{Chen2022} and
\citet{Suresh2025} refer to the same 2004 6dFGS
spectrum; their disagreement is not a temporal
change in IC~1495. The broad-line selection of
\citet{Chen2022} was largely automated, whereas
\citet{Suresh2025} fitted and visually inspected
their selected spectra. \citet{Hon2025} have already
documented substantial disagreements between
automated and visually verified 6dFGS broad-line
catalogues. Our object-level analysis illustrates
why conflicting catalogue types cannot themselves
establish a changing-look event. In NGC~2992,
repeated spectroscopy and a long X-ray record link
broad-line visibility to changes in nuclear luminosity
\citep{Guolo2021NGC2992}. For IC~1495, the
model-dependent 2024 candidate and the limited
2004 recovery do not justify a changing-look
classification.

\subsection{Host structure and gas supply in a 2MIG-selected galaxy}
\label{subsec:discussion_host}

The stellar bar and coherent outer $m=1$ mode establish substantial
non-axisymmetry in this 2MIG-selected galaxy
\citep{Karachentseva2010,Pulatova2015,Durbala2009}.
A stellar $m=1$ mode is compatible with the isolation selection:
such asymmetries can arise through asymmetric gas accretion or
persist after earlier perturbations \citep{Bournaud2005}.

The global H\,I profile establishes an atomic-gas reservoir of
approximately $7\times10^9\,M_\odot$ and is consistent with the
adopted symmetric-profile null
(Fig.~\ref{fig:ic1495_hi_profile}).
The $m=1$ mode describes the spatial stellar distribution, whereas
the H\,I test measures flux balance in an unresolved velocity
profile; their results are therefore compatible.
Bars can drive gas redistribution within galactic discs
\citep{Fanali2015}. The observed bar and atomic-gas reservoir
make inward gas transport a physically plausible context for the
Seyfert nucleus. Spatially resolved gas kinematics across the bar
and inner disc would test whether this pathway operates in IC~1495.

\section{Conclusions}
\label{sec:conclusions}

We reassessed the evidence for broad H$\alpha$ in IC~1495
using the 2004 and 2024 optical spectra, with complementary
observations of its nucleus and 2MIG-selected host.

\begin{enumerate}

\item
The 2024 H$\alpha$+[N~II] blend admits a broad-H$\alpha$
candidate with an intrinsic FWHM near
$2000~\mathrm{km\,s^{-1}}$ in the lowest-BIC fit among the
tested models. Its preference disappears when H$\alpha$ and
[N~II] share two narrow kinematic components, and its
statistical support depends on the noise prescription.
Broad H$\alpha$ is undetected in 2004 under the adopted
criterion, but the specified reference profile is recovered
in only 11--55 per cent of the injection--recovery trials
across the tested grid. Thus, neither the identification of
a BLR component in 2024 nor its absence in 2004 is secure
enough to establish a changing-look transition. Aperture
and calibration differences also limit comparisons of
absolute broad-line flux.

\item
The [O~III] profile favours a broader, blueshifted component,
whereas single-component fits are preferred for [S~II] and
[O~I] at the achieved sensitivity. This ionized-gas
kinematic complexity motivates testing alternative
narrow-line descriptions of the H$\alpha$+[N~II] blend.
The [O~III] asymmetry is consistent with an outflow
interpretation.

\item
Narrow-line excitation, compact absorbed X-ray emission
and long-term optical and MIR variability support an
absorbed, variable Seyfert nucleus independently of the
broad-H$\alpha$ candidate. The aperture-integrated light
curves yield no stable optical--MIR lag, and the IAC80
sequences show no significant intraday variability at
the achieved sensitivity.

\item
The host contains a prominent stellar bar, a coherent outer
$m=1$ mode and an atomic-gas reservoir of approximately
$7\times10^9\,M_\odot$. Its unresolved H\,I profile is
consistent with the adopted symmetric-profile null.
The bar provides a plausible route for gas redistribution
in this non-axisymmetric disc; spatially resolved gas
kinematics would test its connection to nuclear activity.

\end{enumerate}

Repeat optical spectroscopy with matched apertures,
comparable resolution and reliable flux calibration would
test the persistence and variability of the 2024
broad-H$\alpha$ candidate.

\section*{Acknowledgements}

The IAC80 observations were obtained at the Teide Observatory,
Tenerife, Spain, operated by the Instituto de Astrofísica
de Canarias. We thank the support astronomers and night
assistants for conducting the service-mode observations.

We acknowledge the teams responsible for the surveys,
facilities, and archival products used in this work,
including 6dFGS, SDSS-V, SkyMapper, the DESI Legacy Imaging
Surveys, ZTF, WISE/NEOWISE, \textit{Chandra}, and the
Extragalactic Distance Database.
This research also used NASA/IPAC IRSA and NED,
NASA's Astrophysics Data System, DESI DR1,
and HIPASS HICAT.

\section*{Data availability}

The 6dFGS spectrum (g2330477--132908; \texttt{SpecID}=132563),
SDSS-V/BOSS spectrum (\texttt{sdss\_id}=69674331; MJD~60323),
SkyMapper DR4, Legacy Survey DR10, ZTF, WISE/NEOWISE,
Nançay H\,I, and \textit{Chandra}/ACIS data (ObsID~23699)
are available from their public archives.
The H\,I beam-confusion assessment also used NED,
DESI DR1, Legacy Survey DR10, and HIPASS HICAT.

IAC80/CAMELOT2 observations were obtained through a dedicated
service-mode programme. Comparison-star coordinates,
APASS DR9 photometry, and a finding chart are available
in the Zenodo \textit{AGN Reference Fields} dataset
\citep{Izviekova2026AGNfields}, version~1.3
(doi:10.5281/zenodo.23265726), with an interactive field at
\url{https://izviekova-blip.github.io/agn-reference-fields/ic1495.html}.

Reduced photometry, continuum and emission-line fits,
broad-H$\alpha$ injection--recovery outputs,
host-structure and H\,I products, and derived time-series
tables are available from the corresponding author
upon reasonable request.

\bibliographystyle{mnras}
\bibliography{library}

\appendix
\setcounter{secnumdepth}{3}

\section{Observational data and validation}
\label{app:data_validation}

The systemic redshift $z_{\rm sys}=0.021295\pm0.000030$
follows from the Nançay H\,I heliocentric velocity
$V_{\rm hel}=6384\pm9$~km~s$^{-1}$
\citep{Theureau1998HI}, using $z=V_{\rm hel}/c$.
EDD lists the same central velocity.
For the distance scale we use $z_{\rm dist}=0.02119$
in flat $\Lambda$CDM with $H_0=70$~km~s$^{-1}$~Mpc$^{-1}$
and $\Omega_{\rm m}=0.30$.
This gives $D_L=92.23$~Mpc, $D_A=88.44$~Mpc,
and $0.4288$~kpc~arcsec$^{-1}$.

\subsection{Optical spectroscopy}
\label{app:optical_products}

\subsubsection{6dFGS archival spectrum}
\label{app:6df_data}

The 6dFGS product g2330477--132908
(\texttt{SpecID}=132563, \texttt{QUALITY}=4)
contains visual, red, and merged spectra.
The red and visual exposures were 600 and 1200~s at
MJD~53240.62 and 53240.64, respectively, on 2004 August 23.
The headers give $Z=0.021102$ and $Z_{\rm HELIO}=0.021128$.
The fibre diameter is $6.7^{\prime\prime}$ and red-arm
sampling is $2.2013$~\AA~pixel$^{-1}$.
The broad-wavelength continuum analysis uses the merged product. Local
fits use the native arm grids and resolution prescriptions
(Appendix~\ref{app:spectral_methods}).
The headers give no physical flux units.
SkyMapper anchors the reconstructed narrow-line fluxes and
Balmer decrement (Appendix~\ref{app:balmer_methods}).
The red-arm broad-H$\alpha$ fits retain the native relative scale.

\subsubsection{SDSS-V/BOSS DR20 spectrum}
\label{app:sdssv_data}

The APO spectrum (\texttt{sdss\_id}=69674331, MJD~60323,
2024 January 14) is from one 900.07-s exposure through
a $2.0^{\prime\prime}$ fibre. Median guider seeing was
$1.653^{\prime\prime}$. The catalogue position
$(352.6989519^\circ,-13.4854791^\circ)$ is $0.30^{\prime\prime}$
from the adopted nucleus $(352.6988750^\circ,-13.4854444^\circ)$.
The daily product is
\texttt{spec-100187-60323-63050395064467474.fits},
with \texttt{ZWARNING}=0, $z_{\rm pipeline}=0.0211605$,
and pipeline class galaxy/AGN.

The H$\alpha$ fits use the native logarithmic grid,
finite flux and inverse variance, \texttt{IVAR}$>0$,
and \texttt{AND\_MASK}=0.
The ancillary line and stellar checks also reject OR bits
16, 17, 18, 19, 22, 23, 25, 26, 27, and 28.
H$\alpha$ models use the wavelength-dependent LSF and vacuum
wavelengths 6564.61~\AA\ for H$\alpha$ and
6549.86/6585.27~\AA\ for [N~II].

The \texttt{spFluxcalib} and \texttt{spFluxcorr} audit
(exposure 00363917, field 100187, fibre~215) traces the
H$\alpha$-region \texttt{BADFLUXFACTOR} OR bit to blue-camera
samples with zero calibration and inverse variance.
Contributing red-camera samples have positive inverse
variance and lack this flag, so the AND bit is unset.
The multiplicative and additive \texttt{spFluxcorr} terms
are unity and zero, respectively.
This audit identifies the flag origin. Absolute flux
calibration remains a separate uncertainty.
COADD and MJD\_EXP share the exposure and test product consistency.

\subsection{Photometric reference and imaging}
\label{app:photometric_imaging_data}

\subsubsection{SkyMapper photometric reference}
\label{app:skymapper_data}

SkyMapper DR4 $g,r,i$ photometry \citep{Onken2024} uses
\texttt{flags}$<4$, \texttt{nimaflags}$<5$, and the AB zero
point of 3631~Jy. The fiducial aperture diameter is
$6^{\prime\prime}$. Combining the 6- and 8-arcsec measurements
tests an effective $6.7^{\prime\prime}$ aperture matched to the
6dFGS fibre (Appendix~\ref{app:balmer_methods}).

\subsubsection{Legacy Survey imaging}
\label{app:legacy_data}

The DR10 $g,r,z$ cutouts \citep{Dey2019} contain
$700\times700$ pixels at $0.262$~arcsec~pixel$^{-1}$
and span approximately $183^{\prime\prime}$.
Fluxes are in AB nanomaggies per pixel and position angles
are east of north. A common dilated segmentation mask
preserves the target and excludes contaminants and defects.
We subtract a sigma-clipped sky plane fitted outside the main
disc in each band and use sector residuals to test the background.
The $r,z$ images constrain the structure and $g$ tests
wavelength dependence. Integrated photometry uses the same
masks, sky, and geometry (Appendix~\ref{app:host_methods}).

\subsection{Optical and MIR monitoring}
\label{app:monitoring_data}

\subsubsection{ZTF and WISE/NEOWISE selection}
\label{app:variability_data}

We follow the variability analysis of \citet{Izviekova2026Mrk42Mrk493}.
We retrieve ZTF $g,r,i$ catalogue light curves \citep{Bellm2019,Masci2019}
from the IRSA Light Curve Service. We retain one object identifier
(OID) per filter from the field- and CCD-quadrant-specific PSF
photometry. Among OIDs with at least three measurements and median
separation within $0.4^{\prime\prime}$ of the closest eligible OID,
we select the one with the most measurements, breaking ties by
brighter median magnitude. We match seeing by exposure ID and
convert the native ZTF magnitudes to flux densities using the AB
zero point of 3631~Jy.

The fiducial cuts are seeing
$\leq3.0^{\prime\prime}$, $0<\sigma_m\leq0.40$~mag, and separation
$\leq1.5^{\prime\prime}$ where available.
We test a $2.0^{\prime\prime}$ seeing cut.
Global and local clipping use 3.5 and 4.0 times the MAD-based
dispersion, respectively, with a 15-d local window and at least
six neighbours. We average accepted magnitudes nightly with
inverse-variance weights and pair bands within $\Delta t\leq0.35$~d.

WISE/NEOWISE \citep{Wright2010,Mainzer2014} selection requires
finite magnitudes and errors, $\mathrm{S/N}_{W1}>5$,
$\mathrm{S/N}_{W2}>3$, \texttt{qual\_frame}$>0$, and zero
contamination flags in both bands.
Consecutive gaps of at most 10~d define epochs.
Within each epoch we apply a two-MAD-dispersion cut, an error limit
of three times the epoch median, and two-interquartile-range
fences for more than four measurements.
We suppress clipping if fewer than three measurements would remain.
Saved cleaning tables record additional inspected exclusions.
Inverse-variance-weighted mean magnitudes give 20 MIR epochs.
The Vega flux zero points are 309.54 and 171.787~Jy for $W1$ and
$W2$ \citep{Cutri2012WISE}.
We correct Galactic extinction before colour and flux analyses.

\subsubsection{IAC80 aperture photometry}
\label{app:iac80_data}

CAMELOT2 has a scale of $0.322$~arcsec~pixel$^{-1}$.
The 2025 October 13 $R$ and October 14 $V$ sequences
comprise 26 exposures of 250~s and 35 of 300~s,
respectively (\texttt{INSFILTE=R} and
\texttt{INSFILTE=V}). The target and comparison stars
were measured with the same circular aperture in each
frame. We adopt a 5-pixel ($1.61^{\prime\prime}$)
radius for both bands and use a 3-pixel
($0.97^{\prime\prime}$) radius to check the
$R$-band intraday result. 
Sky levels for the initial light curves were estimated
with sigma clipping in 9-pixel-wide annuli beginning
120 pixels beyond each aperture. 
The three numbered local comparison stars, their J2000
coordinates, APASS DR9 photometry, and a static finding
chart are available at
\url{https://izviekova-blip.github.io/agn-reference-fields/ic1495.html}.
The $R$ magnitudes listed there are estimates transformed
from APASS $r'$ and $i'$; the intraday tests use
instrumental differential magnitudes measured from the
CAMELOT2 frames. 

\subsection{\textit{Chandra} processing}
\label{app:chandra_data}

We reprocess the ungrated ACIS-S observation
(ObsID~23699, 2020 December 5, approximately 3.0~ks)
with CIAO~4.18 and \texttt{chandra\_repro}.
\texttt{fluximage} supplies the 0.5--7.0~keV image,
exposure map, and PSF map.
\texttt{wavdetect} uses scales of 1, 2, 4, and 8 pixels and
a threshold of $10^{-6}$.
We compare the X-ray and optical positions.
\texttt{srcflux} uses a PSF-based region, local background, and
\texttt{arfcorr} in the 0.5--7.0, 0.5--2.0, and 2.0--7.0~keV bands.
The fits use ungrouped source and background spectra, an ancillary
response file (ARF), and a redistribution matrix file (RMF).

\subsection{Nançay H\,I beam screening}
\label{app:hi_confusion}

The profile comes from the EDD All Digital H\,I Profile Catalog
\citep{Courtois2009}. For east--west and north--south offsets
$(\Delta E,\Delta N)$, we use an elliptical Gaussian with
$\mathrm{FWHM}_{\rm EW}=3.5^{\prime}$ and
$\mathrm{FWHM}_{\rm NS}=23^{\prime}$, giving the response
\begin{equation}
B=\exp\left[-4\ln2\left(
\frac{\Delta E^2}{\mathrm{FWHM}_{\rm EW}^{2}}+
\frac{\Delta N^2}{\mathrm{FWHM}_{\rm NS}^{2}}
\right)\right].
\label{eq:nancay_beam_response}
\end{equation}
We cross-match NED with Legacy DR10 and consolidate duplicates
by \texttt{LS\_ID}, NED association, and position within
$1.5^{\prime\prime}$. We compare 95-per-cent photometric-redshift
intervals with $z_{\rm sys}$. Secure spectroscopic matches require
$|\Delta v|\leq600$~km~s$^{-1}$, with
\begin{equation}
\Delta v=c\frac{z_{\rm cand}-z_{\rm sys}}{1+z_{\rm sys}}.
\label{eq:confusion_velocity}
\end{equation}
The intervals for seven resolved high-response sources exclude
$z_{\rm sys}$. The unresolved WISEA~J233046.93$-$132352.2
remains a possible contributor because its redshift posterior is
unconstraining. The main-beam response to
2MASX~J23322273$-$1314285 ($z=0.021091$) is negligible at its
$27.38^{\prime}$ separation and large east--west offset.
HIPASS HICAT has no entry within $18^{\prime}$
(Section~\ref{sec:ic1495_hi_results}).

\section{Analysis methods and sensitivity tests}
\label{app:methods}

\subsection{Continuum and narrow-line diagnostics}
\label{app:spectral_methods}

\subsubsection{6dFGS continuum and resolution}
\label{app:6df_continuum_grid}

\begin{figure}
\centering
\includegraphics[width=\columnwidth]{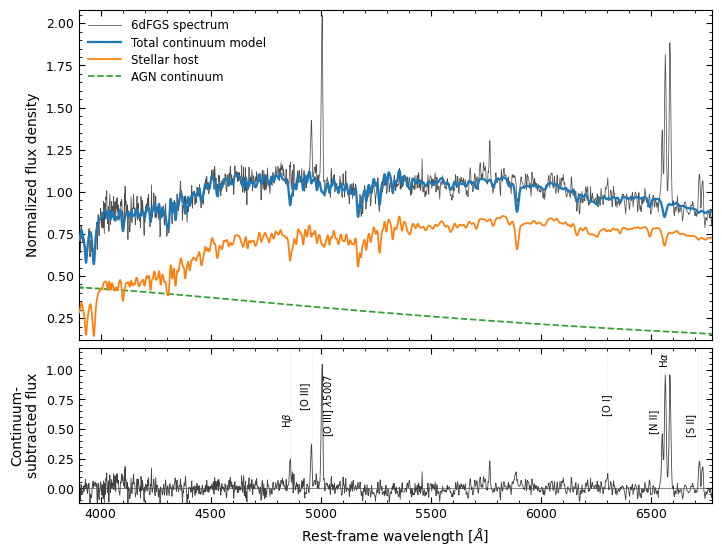}
\caption{Reference 6dFGS host--AGN decomposition. The upper panel
shows the data, total continuum, and stellar and nuclear
contributions. The lower panel shows the continuum-subtracted
spectrum.}
\label{fig:ic1495_continuum}
\end{figure}

We fit E-MILES templates \citep{Vazdekis2016} with
\textsc{ppxf} \citep{Cappellari2017}.
The broad-wavelength model includes a nuclear continuum
$F_\nu\propto\nu^{1/3}$ ($F_\lambda\propto\lambda^{-7/3}$)
with fitted normalisation and reddening.
We omit optical Fe~II from the adopted grid after testing it.
Observed-frame visual/red FWHM pairs $(5.0,9.0)$, $(5.5,10.5)$,
and $(6.0,12.0)$~\AA\ are transformed to the local rest frame,
with a smooth transition across the arm overlap.
Two emission masks and multiplicative degrees $m=0,2,4$ give
18 configurations. The conservative mask widens exclusions
around strong lines and the splice.
We re-optimise stellar weights, shift, effective broadening,
nuclear normalisation, and reddening and compare host-only
and host+AGN fits.
Host fractions use 5080--5120~\AA\ and the combined
6400--6450 and 6650--6690~\AA\ intervals.
Figure~\ref{fig:ic1495_continuum} shows the intermediate
resolution, baseline mask, and $m=2$.

The separate native red-arm grid uses templates with 24 ages
(0.063--12.589~Gyr), six metallicities
($[M/H]=-1.71$ to $+0.22$), and intrinsic FWHM 2.51~\AA.
Fits include a featureless nuisance continuum and
multiplicative degrees $m=2,4$. The continuum-fit masks
exclude the H$\alpha$ complex, other emission features,
and adopted sky-contaminated intervals.
Unweighted fits to saved sky features near 5577, 6300, and 6364~\AA\
in \texttt{SPECTRUM R} use pixels within 22~\AA\ of each nominal
centre, a pixel-integrated Gaussian, and a linear background.
The fitted FWHMs are 7.20, 7.26, and 8.28~\AA, respectively.
We test their rounded extremes (7.2 and 8.3~\AA) as constant
observed-frame LSF scenarios at H$\alpha$, assuming transfer
from shorter wavelengths.

\subsubsection{Local 6dFGS line diagnostics}
\label{app:6df_line_models}

At $z_{\rm 6dF}=0.021102$, fit windows are
4780--5120~\AA\ (H$\beta$+[O~III]),
6250--6400~\AA\ ([O~I]), 6480--6640~\AA\ (H$\alpha$+[N~II]),
and 6660--6770~\AA\ ([S~II]).
Each fit includes a linear residual continuum and instrumental
broadening. The initial fits share kinematics within each complex
and fix $F_{5007}/F_{4959}=2.98$, $F_{6583}/F_{6548}=2.96$,
and $F_{6300}/F_{6363}=3.0$.
The [S~II] ratio is free.
We compare free [O~I] kinematics with ties to H$\alpha$+[N~II]
or [S~II] and test an asymmetric component in H$\beta$+[O~III].
Appendix~\ref{app:injection_recovery} gives the separate
broad-line prescription.

For the 3000 narrow-line Monte Carlo trials we multiply the formal errors by
\begin{equation}
s=\max\left(1,\sqrt{\chi_\nu^2}\right).
\label{eq:error_rescaling}
\end{equation}
This rescaling applies to the narrow-line uncertainty
calculations. The native broad-line simulations in
Appendix~\ref{app:injection_recovery} retain the original
pixel-error vector.
We report 16th--84th percentiles and recompute BPT coordinates
\citep{Baldwin1981} using the boundaries of
\citet{Kauffmann2003,Kewley2001,Kewley2006}.
The [S~II] density uses \textsc{PyNeb} \citep{Luridiana2015}
at $10^4$~K, with sensitivity tests over 8000--15\,000~K.
For [O~I], 3000 null trials compare the doublet with a local
continuum and alternative windows test continuum sensitivity.

\subsubsection{BOSS continuum and WHAN}
\label{app:boss_continuum_whan}

Four no-Fe~II continua combine $m=4,8$ with WDISP/WRESL.
We hold the global continuum fixed in local fits and
distinguish matching and crossed continuum/line LSFs.
WHAN uses the best narrow-only fit in each saved 2004
scenario included in the diagnostic and in each of the two
2024 WDISP configurations.
For two components we sum fluxes separately for H$\alpha$
and [N~II]~$\lambda6583$.
Rest-frame equivalent widths use the global continuum plus
the local background.
Strong-AGN boundaries are
$\log_{10}([\mathrm{N\,II}]/\mathrm{H}\alpha)>-0.4$
and $\mathrm{EW}(\mathrm{H}\alpha)>6$~\AA.
The scenario ranges quantify decomposition sensitivity.

\subsubsection{SkyMapper flux reconstruction}
\label{app:balmer_methods}

We reconstruct line fluxes from the local photometric continuum
$f_{\lambda,{\rm phot}}$ and observed equivalent widths using
\begin{align}
{\rm EW}_{\rm obs}&=(1+z_{\rm 6dF}){\rm EW}_{\rm rest},
\label{eq:ew_observed}\\
F_{\rm line}&={\rm EW}_{\rm obs}\,f_{\lambda,{\rm phot}}.
\label{eq:line_flux_photometric}
\end{align}
Neighbouring bands provide the continua near H$\beta$ and [O~III].
For H$\alpha$ we compare fiducial $r-i$ interpolation with $g-r$
extrapolation. Aperture choice, seeing, and non-simultaneity affect
the reconstruction.
We use the Fitzpatrick law \citep{Fitzpatrick1999}, GNILC
foreground reddening, and SFD as a sensitivity test.
We propagate line and photometric errors through 30\,000
Monte Carlo and bootstrap trials.
For $(\mathrm{H}\alpha/\mathrm{H}\beta)_0=3.10$ and $R_V=3.1$,
\begin{equation}
E(B-V)_{\rm NLR}=
\frac{2.5}{k({\rm H}\beta)-k({\rm H}\alpha)}
\log_{10}\left[
\frac{({\rm H}\alpha/{\rm H}\beta)_{\rm obs}}
{({\rm H}\alpha/{\rm H}\beta)_0}\right].
\label{eq:nlr_ebv}
\end{equation}
Here $k(\lambda)=A_\lambda/E(B-V)$ is the adopted extinction curve.
The [O~III]~$\lambda5007$ comparison with \citet{Suresh2025}
uses foreground-corrected fluxes before internal NLR correction.

\subsection{Gas kinematics and ancillary spectra}
\label{app:gas_profile_methods}

\subsubsection{Independent BOSS narrow-line fits}
\label{app:boss_narrow_kinematics}

We compare one- and two-component fits to [O~III], [O~I],
and [S~II] across four continuum/LSF configurations.
Fits include local linear backgrounds and instrumental broadening.
The [O~III] and [O~I] doublet ratios are fixed and the [S~II] ratio
is free. The score $\Delta{\rm BIC}_{1-2}={\rm BIC}_1-{\rm BIC}_2$
compares the fits at the sensitivity of each line.
The wing sensitivity varies across transitions.
We report the [O~III] wing width, flux fraction, and velocity
relative to the core and [S~II] gas.
The intrinsic profile has cumulative flux fractions of 0.10, 0.50,
and 0.90 at $v_{10}$, $v_{50}$, and $v_{90}$, respectively, giving
$W_{80}=v_{90}-v_{10}$ before instrumental convolution.

Each configuration has 199 parametric-bootstrap trials for
two-component [O~III] and single-component [S~II].
Unscaled diagonal Gaussian errors are independent between doublets.
Standard-normal draws are shared across continuum/LSF configurations.
We refit amplitudes, kinematics, and backgrounds at fixed continuum,
LSF, and component count and audit failures and parameter boundaries.
The conditional 16th--84th and 2.5th--97.5th percentiles describe
the Gaussian-noise experiment. We report the configuration spread
separately.
\textsc{PyNeb} \citep{Luridiana2015} converts valid [S~II] ratios to
single-zone densities at $10^4$~K, with sensitivity tests at
8000, 15\,000, and 20\,000~K and out-of-grid ratios recorded as
limits. These densities describe the [S~II]-emitting gas at fixed
temperature and atomic data.

\subsubsection{Ancillary spectral checks}
\label{app:additional_line_profiles}
\label{app:stellar_kinematic_check}

Local observed-flux tests use the stricter OR masks
(Appendix~\ref{app:sdssv_data}), polynomial backgrounds,
and signed neighbouring-line and absorption proxies.
We distinguish missing wavelength coverage from unsuccessful fits.
The [Ne~III]/[S~III] grid varies WDISP/WRESL, polynomial degrees
1--3, and nuisance dispersions of 150 and 270~km~s$^{-1}$.
The follow-up uses WDISP, a quadratic background, and
150~km~s$^{-1}$.
Target areas are non-negative and profiles include the LSF
and pixel integration.
The [Ne~III] best-fit intrinsic FWHM is $51$~km~s$^{-1}$, with
zero intrinsic width in 103/199 bootstrap trials. The profiled
$\Delta\chi^2=2.71$ width crossing is nominal.
For [S~III] we refit the principal kinematics, ratio, and nuisance
terms with and without a fixed [O~III]-wing template with a
shared doublet ratio.
The one-component fit gives FWHM $305$~km~s$^{-1}$ and
$F_{9531}/F_{9069}=2.63$.
Adding the wing gives $\Delta\chi^2=4.09$ and a conditional
Gaussian tail fraction of 0.030 from 199 single-component null
trials, using the plus-one count.
The test conditions on the external template and omits stellar
mismatch, sky/telluric, and correlated-noise systematics.
The 6382-\AA\ search gives maximum template signal-to-noise ratios
of 1.94--2.08 and Gaussian search tails of 0.31--0.42.

The stellar \textsc{ppxf} check varies 5100--5600 and
4700--5600~\AA\ (rest-frame vacuum), additive degrees 4/8,
and both LSF prescriptions.
We transform E-MILES wavelengths and intrinsic widths from air
to vacuum and apply strict quality, emission, and library-resolution
cuts. The retained pixels exclude Mg~b, Fe5270, and Fe5335.
The fitted broadenings serve as diagnostics.

\subsection{Broad Balmer emission and cross-epoch detectability}
\label{app:broad_balmer_methods}

\subsubsection{SDSS-V H$\alpha$ models}
\label{app:sdssv_halpha_models}

The six reference fits use 159 native pixels in the
6425--6665~\AA\ vacuum rest-frame window, with nominal weights,
a wavelength-dependent LSF, and pixel-integrated profiles.
Three narrow-line prescriptions use one component with shared
H$\alpha$/[N~II] kinematics, independent narrow H$\alpha$/[N~II]
kinematics, or two components with shared kinematics.
Each has a linear background and is fitted with and without a
non-negative broad H$\alpha$ component.
The global continuum is fixed. The broad H$\alpha$ centroid velocity
ranges from $-1000$ to $+1000$~km~s$^{-1}$ relative to the
systemic redshift $z_{\rm sys}=0.021295$.
Its intrinsic FWHM ranges from 1000 to 8000~km~s$^{-1}$.
We test continuum, Fe~II, LSF, and fixed [N~II] ratios
3.05 and 2.96. Equation~\ref{eq:sb_broad} defines $S_{\rm B}$.

The extended grid fixes the ratio at 2.96 and uses four
no-Fe~II continua and local degrees 1/2.
The additional narrow prescriptions are fixed [S~II],
[S~II] plus a fixed [O~III] wing, and [S~II] plus a free
component.
H$\alpha$ and [N~II] areas are independently non-negative
subject to the doublet constraint.
Each $\Delta{\rm BIC}_{\rm narrow-broad}$ compares a pair
of fits within one prescription. Scores for externally fixed
templates are not directly ranked against those for freely
fitted kinematics. Null calibration applies to the reference
six-model search with a local linear background; it does not
extend to the added template or quadratic-background models.

For each of the four configurations, we use 49 consecutive
paired [S~II]/[O~III] bootstrap draws to perturb the fixed
[S~II] and [S~II]+wing templates.
We refit the same H$\alpha$ data with linear local backgrounds
and with and without broad emission.
We retain failed, boundary, and zero-area trials.
The width and centroid are undefined at numerically zero
broad area. The quantiles describe sensitivity to the supplied
kinematics at fixed H$\alpha$ data.

\subsubsection{2024 null calibration}
\label{app:sdssv_empirical_nulls}

Observed and simulated fits use matched multistart optimisation
and transfers between nested models to check minimum-$\chi^2$
ordering at zero broad area.
Strengthened multistart tests at $R_{\rm NII}=2.96$ reuse the
generators and noise draws.
Each null generator supplies 199 trials at fixed fitted
narrow-only parameters, repeating all six fits and the
$S_{\rm B}$ selection.
Gaussian runs use degree-4 WDISP at ratio 3.05 and
degree-4/degree-8 WDISP at ratio 2.96.

The blue control residuals come from 6000--6100 and
6150--6250~\AA, and the red control residuals from
6805--6855~\AA, all in the vacuum rest frame.
The masks exclude 6076--6101~\AA\ in the rest frame and
6294--6310 and 6357--6372~\AA\ in the observed frame.
The two blue windows retain 54 and 49 pixels, and the red
window retains 32 pixels. We divide the continuum residuals by
their nominal errors and subtract the mean within each window.
Resampling blocks stay within contiguous segments of retained
pixels. We centre each assembled residual pool without rescaling
its variance and retain the nominal fit weights.
For the degree-4 WDISP continuum, the detrended residual
RMS in units of nominal pixel uncertainties is 0.59 and
0.70 in the two blue control windows and 0.88 in the red
control window. The corresponding five-pixel block pools
have RMS 0.64 for the combined blue controls and
0.86 for the red control. The control-residual autocorrelation was also measured
at lags of 1--12 native pixels. Adjacent-pixel
correlation is stronger in the red control window
than in either blue window.

For degree-4 WDISP at $R_{\rm NII}=2.96$, the lowest-BIC
fits without and with broad H$\alpha$ have nominal
$\chi^2/(n-q)=0.297$ and 0.237, respectively
(Table~\ref{tab:ic1495_ha_free_model_bic}).
For these same two fit prescriptions, 199 realisations
of the two-component narrow-only generator give median
pairs of $(0.98,1.00)$ for nominal Gaussian noise,
$(0.41,0.44)$ for blue-control blocks, and $(0.68,0.71)$
for red-control blocks.
The observed values lie below the corresponding medians.
These residual-scale diagnostics leave the adequacy
of the transferred control-window noise uncertain.

At ratio 3.05, we use 3-, 5-, and 8-pixel blocks from the blue
controls and 5-pixel blocks from the red control.
At ratio 2.96, two WDISP degrees, two five-pixel residual pools,
and three generators give 2388 trials (199 per cell).
The calibration assumes that each control residual pool
represents noise in the line window; $S_{\rm B}$ is calibrated
separately for the Gaussian and empirical noise prescriptions.
We use the plus-one tail count in
equation~\ref{eq:conditional_null_tail} and report binomial
Monte Carlo intervals separately from model sensitivity.
The maximum over the three fitted generators describes
conditional tail probabilities for the tested nulls.

\subsubsection{Conditional broad-H$\beta$ tests}
\label{app:sdssv_hbeta_models}

Fits use 143 native pixels at 4770--4930~\AA\ in the vacuum
rest frame. Broad H$\beta$ has non-negative area and adopts
the reference-H$\alpha$ FWHM of 1928.90~km~s$^{-1}$ and
$v=68.67$~km~s$^{-1}$.
We refit the narrow lines and linear/quadratic backgrounds
across four fixed continua (degrees 4/8, WDISP/WRESL), using
the reference WDISP line LSF, common mask, and nominal errors.
Profiling the broad area gives the nominal
$\Delta\chi^2=2.71$ endpoint at fixed H$\alpha$ reference.
The conditional Balmer-ratio endpoint has uncalibrated
coverage, so we leave BLR extinction undetermined.
Null tests use 199 trials per local degree and noise
prescription (Gaussian and five-pixel blocks).
Injections use 99 trials per configuration at
broad H$\alpha$/H$\beta=3.1,5,10$ with conditional
null-calibrated $\alpha=0.05$.

\subsubsection{2004 null calibration and injection--recovery}
\label{app:injection_recovery}
\label{app:injection_sensitivity}

The eight configurations combine observed-frame LSF FWHM
7.2/8.3~\AA, global multiplicative degree $m=2,4$, and local
additive degree $p=1,2$. The baseline uses 7.2~\AA, $m=2$, $p=1$.
All fits retain 79 native red-arm pixels, the header wavelength
solution, and pixel boundaries.
Air wavelengths are 6562.801~\AA\ for H$\alpha$ and
6548.05/6583.45~\AA\ for [N~II], with $z_{\rm ref}=0.02113$
and fixed [N~II] ratio 3.05.
We integrate Gaussian profiles over pixels after convolution
with a Gaussian LSF.

The three original narrow-line prescriptions are fitted with
and without broad H$\alpha$.
Narrow velocities span $-500$ to $+800$~km~s$^{-1}$ and
intrinsic dispersions 0--450~km~s$^{-1}$.
Broad centroids span $-1000$ to $+1500$~km~s$^{-1}$ and
intrinsic FWHM 800--6000~km~s$^{-1}$.
Line areas are non-negative and convolution is retained at
zero intrinsic width.
For fixed kinematics, amplitudes and local polynomials are
solved jointly. Multiple starts explore nonlinear parameters.
At fixed global continuum,
\begin{equation}
{\rm BIC}=\chi^2+k\ln n,
\label{eq:bic_native}
\end{equation}
where $n=79$ and $k$ counts fitted line and local-background
parameters.
The reference component has intrinsic FWHM 1928.90~km~s$^{-1}$
and relative flux
\begin{equation}
r_{\rm ref}=F_{\rm broad}/F_{\rm narrow}=0.821416.
\label{eq:reference_injection_ratio}
\end{equation}
The baseline configuration additionally tests
$0.5r_{\rm ref}$ and $1.5r_{\rm ref}$ at the same intrinsic
width. Across all eight configurations, the non-zero
injection uses $r_{\rm ref}$.

Gaussian simulations use unscaled native errors.
The 474 continuum-fit pixels outside the H$\alpha$ window supply
438 overlapping five-pixel blocks of standardised residuals per
continuum configuration. Blocks remain within contiguous runs.
We centre the pool and preserve its variance, then multiply the
resampled residuals by line-window errors.
For each configuration and noise prescription, each fitted
narrow-only generator supplies 199 null trials with the same
six-model search. For generator $g$,
\begin{equation}
p_g(S)=
\frac{1+\sum_{j=1}^{N_{\rm null}}
\mathbb{I}(S_{{\rm B},g,j}\geq S)}{N_{\rm null}+1},
\qquad N_{\rm null}=199,
\label{eq:conditional_null_tail}
\end{equation}
where $\mathbb{I}$ is the indicator function.
The operational criterion is
\begin{equation}
S_{\rm B}>0
\quad\mathrm{and}\quad
p_{\max}=\max_g p_g(S_{\rm B})\leq0.05.
\label{eq:recovery_criterion}
\end{equation}

The injected area scales with the generator's narrow H$\alpha$
flux and the centroid follows the narrow velocity.
For two components we use the summed flux and flux-weighted centroid.
Each generator, noise prescription, and injection level has
99 trials. Zero-injection trials are independent of the
null-calibration samples and estimate the conditional
false-positive fraction.
All six fits are repeated with broad area, width, and centroid free.
The eight-configuration reference grid contains 48 null
ensembles and 96 reference-injection or zero-injection cells. We audit scores, counts, and boundaries.
Wilson intervals give binomial uncertainty at fixed continuum,
LSF, errors, and estimated null distribution.
The between-configuration spread measures sensitivity to modelling
choices and the fractions are conditional on the injected profile.

\subsection{X-ray fits and line search}
\label{app:xray_methods}

We fit the ungrouped spectra and responses in Sherpa over
0.5--7.0~keV using the Poisson W statistic, \citet{Wilms2000}
abundances, and \citet{Verner1996} cross-sections.
We fix $N_{\rm H,Gal}=2.23\times10^{20}$~cm$^{-2}$ and use
$z=0.02119$ for the intrinsic absorber.
We compare a Galactic-only power law with
\begin{equation}
{\tt TBabs}_{\rm Gal}\times{\tt zTBabs}_{\rm int}
\times{\tt powerlaw}.
\label{eq:xray_absorbed_powerlaw}
\end{equation}
We profile column density and photon index jointly and use
$\Gamma=1.7,1.8,1.9$ in fixed-slope tests, with 1.8 fiducial.
Column intervals use 90-per-cent profile likelihoods.
The rest-frame 2--10~keV luminosities include extrapolation
beyond the observed 7-keV fitting limit.

We model neutral Fe~K$\alpha$ with a redshifted Gaussian at
6.40~keV and intrinsic $\sigma=0.01$~keV, testing
\begin{gather}
{\tt TBabs}_{\rm Gal}\times{\tt zTBabs}_{\rm int}
\times({\tt powerlaw}+{\tt zgauss}),
\label{eq:xray_fek_absorbed}\\
{\tt TBabs}_{\rm Gal}\times
[{\tt zTBabs}_{\rm int}\times{\tt powerlaw}+{\tt zgauss}].
\label{eq:xray_fek_external}
\end{gather}
We profile the line normalisation while refitting the continuum.
The 90-per-cent endpoint uses $\Delta W=2.706$ for one parameter.
Equivalent widths refer to the fitted continuum at the rest-frame
line energy.

\subsection{Photometric variability}
\label{app:variability_methods}

\subsubsection{Colour, flux, and lag diagnostics}
\label{app:chromatic_variability_methods}

Colour--magnitude fits use York regression \citep{York2004}
with covariance from the shared magnitude.
Flux--flux fits use orthogonal-distance regression,
\begin{equation}
F_{\nu,1}=aF_{\nu,2}+b.
\label{eq:flux_flux}
\end{equation}
For $a>0$, the effective index in $F_\nu\propto\nu^{-\alpha_{\rm var}}$
and the corresponding AB and Vega colours are
\begin{gather}
\alpha_{\rm var}=-\frac{\ln a}{\ln(\nu_1/\nu_2)},
\label{eq:alpha_var}\\
(m_1-m_2)_{\rm var}=-2.5\log_{10}a,
\label{eq:variable_colour_ab}\\
(W1-W2)_{\rm var}=-2.5\log_{10}
\left[a\frac{F_{\nu,0}(W2)}{F_{\nu,0}(W1)}\right].
\label{eq:variable_colour_wise}
\end{gather}
The slope describes correlated aperture variability and constant
flux contributes to the intercept.
Several variable components, delays, and non-linear responses
can affect the inferred colours.

The bidirectional interpolated cross-correlation function (ICCF)
uses nightly ZTF $g$ fluxes and 11 MIR epochs within the optical
coverage.
The conditional accretion-disc correction extrapolates the
optical aperture flux, which includes starlight, as
$F_\nu\propto\nu^{s_{\rm AD}}$ with fiducial $s_{\rm AD}=1/3$
and alternatives 0.1 and 0.
The observer-frame grid spans $-200\leq\tau\leq250$~d in 1-d
steps, with $\tau>0$ for an MIR delay.
We search $0\leq\tau\leq220$~d for the primary peak,
requiring at least eight overlapping pairs per contributing direction.
The centroid uses adjacent retained lags with $r\geq0.8r_{\rm max}$,
at least two samples, and a positive sum of correlation weights.

Each of the 5000 flux-randomisation and random-subset selection
(FR/RSS) trials per band resamples both light curves with
replacement. An epoch drawn $n$ times is retained once and its
flux is perturbed with Gaussian error $\sigma/\sqrt{n}$.
The ICCF requires at least eight distinct epochs per light curve.
For 11 draws from 11 MIR epochs, this occurs with 36.6 per cent
probability in the MIR. Each band yields 1823 finite peaks.
We obtain 1431 finite centroids for $W1$ and 1458 for $W2$.
We require a distinct correlation feature and a stable, unimodal
centroid distribution to adopt a lag. For IC~1495 the conditional
16th--84th percentile ranges are 42--204~d for $W1$ and 41--192~d
for $W2$ in the observer frame. The broad, multimodal distributions
and weak nominal correlations leave the reverberation lag and
radius undetermined.

\subsubsection{Intraday variability}
\label{app:iac80_variability_methods}

We form differential instrumental-magnitude light curves
between IC~1495 and reference star~2 and between each of
comparison stars~1 and 3 and the same reference. Star~2 was
selected because its median measured magnitude was closest
to that of IC~1495 in the fiducial 5-pixel data; it was kept
fixed for the aperture comparison.
Following \citet{deDiego2014}, the power-enhanced $F$
statistic is
\begin{equation}
F_{\rm enh}=\frac{s_{\rm target-ref}^2}{s_c^2},
\label{eq:f_enh}
\end{equation}
where $s_{\rm target-ref}^2$ is the sample variance of the
target--reference curve. To calculate $s_c^2$, we subtract
the mean from each comparison--reference curve and scale
its squared deviations by the ratio of the mean squared
target--reference uncertainty to the mean squared
uncertainty of that comparison--reference curve. We add
the two scaled sums and divide by $2(N-1)$.
For $N$ measurements in each curve, the nominal $F$
distribution has $\nu_1=N-1$ and $\nu_2=2(N-1)$
degrees of freedom. We classify a sequence as variable
when $F_{\rm enh}\geq F_{\rm crit}$ at $\alpha=0.01$;
the reported $p$-values are upper-tail probabilities.
The aperture radii and photometric extraction are
described in Appendix~\ref{app:iac80_data}.

\subsection{Host photometry and structure}
\label{app:host_methods}

\subsubsection{Disc geometry and stellar mass}
\label{app:host_photometry_methods}

We use contour-based ellipses and \texttt{photutils.isophote}
\citep{Jedrzejewski1987}. The outer-disc selection requires a
populated contiguous interval with $\sigma_q\lesssim0.025$,
$\sigma_{\rm PA}\lesssim2^\circ$, and weak radial trends,
followed by inspection.
Inclination follows
\begin{equation}
\cos^2 i=\frac{q^2-q_0^2}{1-q_0^2},
\label{eq:disc_inclination}
\end{equation}
where $q$ is the projected outer-disc axis ratio and $q_0$ is the
assumed intrinsic disc thickness, with $0.13\leq q_0\leq0.25$.
Elliptical curves of growth use $2^{\prime\prime}$ annuli to
$68^{\prime\prime}$ and fiducial $R_{\rm phot}=64^{\prime\prime}$.
We compare annular-median filling with area correction and
estimate the outer light from an exponential fitted over
$44^{\prime\prime}\leq R\leq60^{\prime\prime}$.
Sensitivity tests vary geometry, centre, sky, and mask dilation.

We transform integrated fluxes to rest-frame SDSS $g,r$ with
\textsc{kcorrect} \citep{BlantonRoweis2007}.
Stellar-mass simulations propagate colour uncertainty, 0.02-mag
calibration, 0.03-mag residual $K$ correction, 0.10-dex scatter
in fiducial $M/L$, and a 300-km~s$^{-1}$ peculiar-velocity term.
We use the BC03/Chabrier and FSPS calibrations of
\citet{RoedigerCourteau2015}, with BC03 as the reference,
and the \citet{Bell2003} relation converted to a Chabrier IMF
\citep{Bernardi2010}. In BC03, Bell, and FSPS order,
$M_\star/L_r$ in solar units follows
\begin{align}
\log_{10}(M_\star/L_r)&=-0.792+1.629(g-r)_{\rm rest},
\label{eq:ml_rc15_bc03}\\
\log_{10}(M_\star/L_r)&=-0.406+1.097(g-r)_{\rm rest},
\label{eq:ml_bell03}\\
\log_{10}(M_\star/L_r)&=-0.647+1.497(g-r)_{\rm rest}.
\label{eq:ml_rc15_fsps}
\end{align}
Excision within $3^{\prime\prime}$ tests the combined effect of
bulge and unresolved nuclear light.

\subsubsection{Fourier diagnostics}
\label{app:fourier_tests}

We expand deprojected intensity as
\begin{equation}
I(R,\phi)=A_0(R)+
\sum_m[a_m(R)\cos(m\phi)+b_m(R)\sin(m\phi)],
\label{eq:fourier_intensity}
\end{equation}
with
\begin{equation}
\frac{A_m}{A_0}=\frac{\sqrt{a_m^2+b_m^2}}{A_0},
\qquad
\phi_m=\frac{\operatorname{atan2}(b_m,a_m)}{m}.
\label{eq:fourier_amplitude_phase}
\end{equation}
The bar radius is bracketed by the outer $m=2$ phase-coherence
boundary and the declining-side half-maximum radius.
We fit the post-bar scale over several intervals in $r,z$,
separately from the aperture correction.
The outer $m=1$ test spans about $40.2^{\prime\prime}$--$51^{\prime\prime}$.
We vary geometry, centre, sky, mask dilation, and radial interval.
Catalogue and inner $r+z$ centres define the principal envelope.
The faintest-isophote centre is an additional test.
We use the circular resultant for phase coherence and check
$r,z$ agreement before combining the bands.
The null scrambles sector fluxes within annuli, preserving
their radial profile and usable-sector counts.
For $k$ exceedances in $N=5000$ trials,
\begin{equation}
p_{\rm MC}=\frac{k+1}{N+1}.
\label{eq:finite_mc_probability}
\end{equation}
The spatial jackknife removes each of eight $45^\circ$ sectors
in turn.

\subsubsection{Two-dimensional fit}
\label{app:imfit_methods}

PSF-convolved \textsc{PyImfit} fits use the sky-subtracted $z$
image, validated mask, and empirical PSF.
The five-component model M4 contains a S\'ersic bulge, Ferrers bar,
exponential disc, point source, and Gaussian ring.
BIC comparisons share region, mask, and error prescription.
We test initialisations and Ferrers bounds and inspect residuals
and boundary flags. Outer isophotes define the deprojection.
The Fourier bracket gives the adopted bar radius.

\subsection{Global H\,I and dynamical quantities}
\label{app:hi_methods}
\label{app:hi_confusion_methods}

For fluxes $F_1,F_2$ on either side of the fitted profile centre,
\begin{equation}
A_{\rm signed}=\frac{F_1-F_2}{F_1+F_2}.
\label{eq:hi_asymmetry}
\end{equation}
The sign indicates the dominant side and the symmetry test uses
$|A_{\rm signed}|$. We add correlated noise to a symmetrised
observed profile and refit the centre and asymmetry.
Noise simulations match the line-free RMS and short-range channel
correlations. We test 225 window, baseline, and symmetric-template
smoothing combinations.

The H\,I mass is
\begin{equation}
M_{\rm HI}=2.356\times10^{5}\frac{D_L^2}{1+z}
\left(\int S_\nu\,dv\right)M_\odot.
\label{eq:hi_mass}
\end{equation}
Here $D_L$ is in Mpc and the integrated flux is in Jy~km~s$^{-1}$.
The uncertainty combines correlated noise, 12.5-per-cent
calibration, and a 300-km~s$^{-1}$ peculiar-velocity dispersion
($300/6384=0.047$, or 4.33~Mpc at $D_L=92.23$~Mpc).
A common distance draw propagates covariance into
$M_{\rm HI}/M_\star$.

For the dynamical check we use the unresolved linewidth proxy
and the stellar-plus-atomic-gas mass including helium.
The mass budget is restricted to these components.
\begin{align}
V_{\rm rot,W}&=\frac{W_{mx}}{2\sin i},
\label{eq:vrot_w}\\
M_{\rm bar,atomic}&=M_\star+1.33M_{\rm HI}.
\label{eq:mbar_atomic}
\end{align}
We compare with the baryonic Tully--Fisher relation (BTFR)
of \citet{McGaugh2012},
\begin{equation}
M_{\rm bar}=A V_f^4,
\label{eq:btfr_reference}
\end{equation}
using $A=47\pm6\,M_\odot({\rm km\,s^{-1}})^{-4}$
and $V_{\rm rot,W}$ as a proxy for the flat rotation speed $V_f$
\citep{Lelli2019}.
We propagate coefficient, mass, inclination, and linewidth
uncertainties jointly.

\section{Supplementary results}
\label{app:supplementary_results}

\subsection{Continuum and narrow-line sensitivity}
\label{app:optical_results_robustness}

Host fractions are stable across the 18 continuum prescriptions.
The fitted nuclear reddening and effective stellar broadening are
sensitive to the continuum shape and LSF, respectively.
All accepted [O~I] prescriptions retain Seyfert excitation.
Using $g-r$ instead of $r-i$ for the H$\alpha$ continuum raises
the median $E(B-V)_{\rm NLR}$ by 0.051~mag.
The reconstructed [O~III] luminosity differs from
\citet{Suresh2025} by 0.026~dex.

\begin{table}
\centering
\caption{Supplementary 2004 fits and photometric reconstruction.}
\label{tab:ic1495_optical_robustness}
\footnotesize
\setlength{\tabcolsep}{3pt}
\renewcommand{\arraystretch}{1.10}
\begin{tabular}{@{}p{0.57\columnwidth}p{0.35\columnwidth}@{}}
\hline
Quantity & Result \\
\hline
Continuum $E(B-V)_{\rm AGN}$ & 0--0.42 mag \\
Models with zero AGN reddening & 12/18 \\
Effective stellar broadening & 101--208 km~s$^{-1}$ \\
$\log([\mathrm{O\,I}]/\mathrm{H}\alpha)$ & $-1.090$ to $-0.914$ \\
Uncalibrated H$\alpha$/H$\beta$ & $5.51^{+0.63}_{-0.52}$ \\
$E(B-V)_{\rm NLR}$ from $g-r$ & $0.767^{+0.102}_{-0.094}$ mag \\
Median $n_{\rm e}$, 8000--15\,000 K & 226--278 cm$^{-3}$ \\
Reconstructed [O~III] flux & $1.50\times10^{-14}$ \\
{[O~III]} aperture change, 6--6.7 arcsec & 8.3 per cent \\
\hline
\end{tabular}
\par\smallskip
\begin{minipage}{\columnwidth}
\footnotesize
\raggedright
\textit{Notes.} Ranges show prescription sensitivity.
The [O~III] flux is foreground-corrected and is in
erg~s$^{-1}$~cm$^{-2}$.
Reference values are in Table~\ref{tab:ic1495_optical_summary}.
The uncalibrated Balmer ratio uses the merged spectrum.
\end{minipage}
\end{table}

\subsection{Broad-H$\alpha$ models and detectability}
\label{app:broadline_results_robustness}

Table~\ref{tab:ic1495_broadline_robustness} gives the full 2004 grid
shown in Fig.~\ref{fig:ic1495_broadline_2004_2024}.
None of the observed fits satisfies
equation~\ref{eq:recovery_criterion}.
Reference recovery spans 11/99--54/99, with 27/99--35/99
for the baseline.
The two-component generator gives 10/99 zero-injection detections
(95-per-cent Wilson interval 0.056--0.176) for block noise at
$(L,m,p)=(8.3,4,1)$ and Gaussian noise at $(8.3,4,2)$.
These are finite-cell estimates under the tested generators.
Calibration is conditional on the three fitted generators.

\begin{table}
\centering
\caption{Native-grid 2004 tests.}
\label{tab:ic1495_broadline_robustness}
\footnotesize
\setlength{\tabcolsep}{2.5pt}
\renewcommand{\arraystretch}{1.10}
\begin{tabular*}{\columnwidth}{@{\extracolsep{\fill}}cccccccc@{}}
\hline
$L$ (\AA) & $m$ & $p$ & $S_{\rm B}$ &
$p_{\max,\rm G}$ & $p_{\max,\rm B}$ & $k_{\rm ref}$ & $k_0$ \\
\hline
7.2 & 2 & 1 &  0.260 & 0.660 & 0.655 & 27--35 & 0--7 \\
7.2 & 2 & 2 & -2.824 & 0.740 & 0.700 & 18--35 & 0--6 \\
7.2 & 4 & 1 &  1.153 & 0.630 & 0.655 & 20--36 & 0--4 \\
7.2 & 4 & 2 & -0.370 & 0.695 & 0.690 & 11--27 & 0--4 \\
8.3 & 2 & 1 &  0.152 & 0.570 & 0.520 & 25--48 & 0--2 \\
8.3 & 2 & 2 & -2.923 & 0.595 & 0.635 & 20--51 & 0--7 \\
8.3 & 4 & 1 &  1.649 & 0.610 & 0.605 & 31--54 & 0--10 \\
8.3 & 4 & 2 & -0.492 & 0.550 & 0.580 & 20--44 & 0--10 \\
\hline
\end{tabular*}
\par\smallskip
\begin{minipage}{\columnwidth}
\footnotesize
\raggedright
\textit{Notes.} $L$ is observed-frame LSF FWHM and $m,p$ are
global multiplicative and local additive degrees.
$p_{\max,\rm G}$ and $p_{\max,\rm B}$ are conditional maxima
over three generators for Gaussian and five-pixel block noise
(199 trials per generator).
$k_{\rm ref},k_0$ are count ranges over generators and noise
prescriptions (99 trials per cell).
The first row is the baseline.
\end{minipage}
\end{table}

At 2024 degree-4 WDISP, linear background, and $R_{\rm NII}=3.05$,
COADD and MJD\_EXP give $S_{\rm B}=5.0979$ and 5.0911.
Both favour two tied components among narrow-only fits and
independent narrow H$\alpha$/[N~II] among broad-containing fits.

The aperture scenarios assume an unresolved BLR, a circular
Gaussian NLR, 6.7/2.0-arcsec fibres, and seeing of 2.0 arcsec
(assumed for 2004) and 1.653 arcsec (measured for 2024).
For NLR FWHM $(0,1,2,3,4,6)$~arcsec, the relative ratio
$r_{2004}/r_{2024}$ is $(1.000,0.824,0.540,0.363,0.273,0.200)$.
These factors illustrate geometric sensitivity.
The eight-configuration recovery grid uses the single
reference ratio in equation~\ref{eq:reference_injection_ratio}.

The narrow-only alternatives in
Table~\ref{tab:ic1495_ha_narrow_components} show how the additional
kinematic structure can account for part of the H$\alpha$ blend.

\begin{table}
\centering
\caption{Narrow-only solutions for the panels of
Fig.~\ref{fig:ic1495_halpha_models}.}
\label{tab:ic1495_ha_narrow_components}
\footnotesize
\setlength{\tabcolsep}{1.2pt}
\renewcommand{\arraystretch}{1.15}
\begin{tabular*}{\columnwidth}{@{\extracolsep{\fill}}lrrrrrrr@{}}
\hline
\shortstack{Panel /\\component} &
$v_{\rm H\alpha}$ &
\shortstack{FWHM\\H$\alpha$} &
$v_{\rm NII}$ &
\shortstack{FWHM\\$[\mathrm{N\,II}]$} &
$f_{\rm H\alpha}$ &
$f_{6583}$ &
$\frac{F_{6583}}{F_{\rm H\alpha}}$ \\
\hline
(a)/1 & $-56.3$ & 221.9 & $-56.3$ & 221.9 & 1.000 & 1.000 & 1.187 \\
(b)/1 & $-74.1$ & 298.5 & $-54.9$ & 281.1 & 1.000 & 1.000 & 1.151 \\
(c)/1 & $-56.3$ & 221.9 & $-56.3$ & 221.9 & 0.643 & 0.728 & 1.276 \\
(c)/2 & $-147.5$ & 637.8 & $-147.5$ & 637.8 & 0.357 & 0.272 & 0.859 \\
(d)/1 & $-59.9$ & 213.4 & $-59.9$ & 213.4 & 0.587 & 0.677 & 1.304 \\
(d)/2 & $-92.8$ & 584.5 & $-92.8$ & 584.5 & 0.413 & 0.323 & 0.886 \\
\hline
\end{tabular*}
\par\smallskip
\begin{minipage}{\columnwidth}
\footnotesize
\raggedright
\textit{Notes.} All solutions use degree-4 WDISP, a local linear
background, and $F_{6583}/F_{6548}=2.96$.
Values are conditional point estimates.
Components 1 and 2 denote the sole or narrower
component and the broader component, respectively.
Centroid velocities use the optical fitting frame, and
intrinsic FWHMs and velocities are in km~s$^{-1}$.
$f_{\ell,i}=F_{\ell,i}/\sum_jF_{\ell,j}$ for line $\ell$
and component $i$.
The final column gives the within-component ratio for the
stronger [N~II] line.
All line areas are positive. The free nonlinear parameters are
interior at a relative boundary tolerance of $10^{-4}$.
\end{minipage}
\end{table}

Table~\ref{tab:ic1495_ha_free_model_bic} gives the six-model
comparison for the configuration of Fig.~\ref{fig:ic1495_halpha_models}.
The lowest BIC is obtained with broad H$\alpha$ and independent
narrow H$\alpha$/[N~II] kinematics. Its BIC is 4.05 below that of
the best narrow-only solution, giving $S_{\rm B}=4.05$.
Within the prescription with two shared narrow components,
adding broad H$\alpha$ increases BIC by 4.43.

\begin{table}
\centering
\caption{Comparison of six fits to the 2024 H$\alpha$+[N~II] blend
with freely fitted narrow-line kinematics.}
\label{tab:ic1495_ha_free_model_bic}
\footnotesize
\setlength{\tabcolsep}{3pt}
\begin{tabular}{@{}lcrrr@{}}
\hline
Narrow-line kinematics & Broad H$\alpha$ & $\chi^2$ & $q$ &
$\mathrm{BIC}-\mathrm{BIC}_{\min}$ \\
\hline
One shared component          & No  & 70.26 &  6 &  9.86 \\
One shared component          & Yes & 48.20 &  9 &  3.01 \\
Independent H$\alpha$/[N~II]  & No  & 61.38 &  8 & 11.11 \\
Independent H$\alpha$/[N~II]  & Yes & 35.06 & 11 &  0.00 \\
Two shared components         & No  & 44.18 & 10 &  4.05 \\
Two shared components         & Yes & 33.40 & 13 &  8.48 \\
\hline
\end{tabular}
\par\smallskip
\begin{minipage}{\columnwidth}
\footnotesize
\raggedright
\textit{Notes.} All fits use the same 159 pixels, degree-4 WDISP
continuum without Fe~II, WDISP line-spread function, linear local
background, and $R_{\rm NII}=2.96$ as in
Fig.~\ref{fig:ic1495_halpha_models}.
The relative BIC uses the minimum over these six fits.
The parameter count $q$ includes the line and local-background
parameters.
\end{minipage}
\end{table}

\subsection{X-ray spectral sensitivity}
\label{app:xray_results_robustness}

\begin{figure}
\centering
\includegraphics[width=\columnwidth]{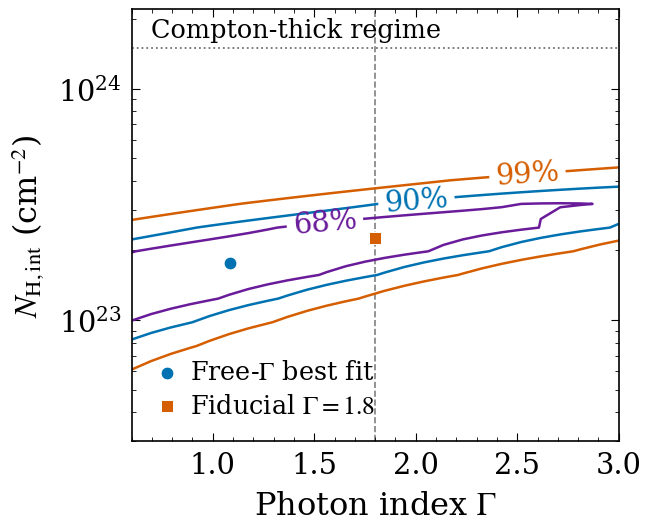}
\caption{Intrinsic column-density and photon-index covariance
for the \textit{Chandra}/ACIS spectrum.}
\label{fig:ic1495_nh_gamma}
\end{figure}

Figure~\ref{fig:ic1495_nh_gamma} shows the column-density and
photon-index covariance, and Table~\ref{tab:ic1495_xray_robustness}
compares fits.
Neutral Fe~K$\alpha$ gives $\Delta W=1.93$.
The 90-per-cent equivalent-width upper endpoints are approximately
1.03 and 0.81~keV
for the absorbed and external-line geometries
(equations~\ref{eq:xray_fek_absorbed} and~\ref{eq:xray_fek_external}).

\begin{table}
\centering
\caption{X-ray fits.}
\label{tab:ic1495_xray_robustness}
\footnotesize
\setlength{\tabcolsep}{3pt}
\renewcommand{\arraystretch}{1.10}
\begin{tabular}{lcccc}
\hline
Model & $\Gamma$ & $N_{\rm H,int}$ & $W$ &
$\log L_{2-10}^{\rm int}$ \\
& & ($10^{23}$ cm$^{-2}$) & & (erg~s$^{-1}$) \\
\hline
Galactic & $-1.62$ & -- & 240.31 & -- \\
Free $\Gamma$ & 1.08 & 1.76 & 232.20 & -- \\
Fixed $\Gamma$ & 1.7 & $2.19^{+0.69}_{-0.56}$ & 232.48 & 42.44 \\
Fixed $\Gamma$ & 1.8 & $2.25^{+0.70}_{-0.57}$ & 232.58 & 42.45 \\
Fixed $\Gamma$ & 1.9 & $2.32^{+0.71}_{-0.57}$ & 232.68 & 42.47 \\
\hline
\end{tabular}
\par\smallskip
\begin{minipage}{\columnwidth}
\footnotesize
\raggedright
\textit{Notes.} The free- and fixed-$\Gamma$ models include
intrinsic absorption.
Column-density intervals use 90-per-cent profile likelihoods.
The fiducial fit has $\Gamma=1.8$.
\end{minipage}
\end{table}

\subsection{Variability and intraday monitoring}
\label{app:time_domain_results}

\subsubsection{Chromatic variability}
\label{app:variability_results_robustness}

The $g-i$ and $r-i$ measurements follow the optical
bluer-when-brighter trend with lower precision
(Table~\ref{tab:ic1495_chromatic_full}).
Figure~\ref{fig:ic1495_chromatic} shows the main flux--flux fits.

\begin{table}
\centering
\caption{Chromatic diagnostics for $N$ paired measurements.}
\label{tab:ic1495_chromatic_full}
\footnotesize
\setlength{\tabcolsep}{1.8pt}
\renewcommand{\arraystretch}{1.12}
\begin{tabular*}{\columnwidth}{@{\extracolsep{\fill}}lccccc@{}}
\hline
\shortstack{Band\\pair} & $N$ &
\shortstack{Colour--\\magnitude slope} &
\shortstack{Flux--flux\\slope $a$} &
\shortstack{Variable\\colour} &
$\alpha_{\rm var}$ \\
\hline
$g-r$ & 186 & $0.798\pm0.082$ & $0.341\pm0.043$ & 1.167 & 3.66 \\
$g-i$ & 43 & $0.747\pm0.206$ & $0.153\pm0.051$ & 2.035 & 3.75 \\
$r-i$ & 51 & $0.674\pm0.184$ & $0.279\pm0.079$ & 1.385 & 6.17 \\
$W1-W2$ & 20 & $-0.470\pm0.098$ & $0.615\pm0.031$ & 1.168 & 1.54 \\
\hline
\end{tabular*}
\par\smallskip
\begin{minipage}{\columnwidth}
\footnotesize
\raggedright
\textit{Notes.} For bands $1-2$, the slopes are for $m_1-m_2$
versus $m_1$ and $F_{\nu,1}$ versus $F_{\nu,2}$, respectively.
Variable colours use AB magnitudes in the optical and Vega
magnitudes for $W1-W2$.
The effective index $\alpha_{\rm var}$ describes correlated
aperture variability
(Appendix~\ref{app:chromatic_variability_methods}).
\end{minipage}
\end{table}

\subsubsection{IAC80 intraday tests}
\label{app:ic1495_idv}

All three sequences yield $F_{\rm enh}<F_{\rm crit}$
at $\alpha=0.01$ (Table~\ref{tab:ic1495_idv_results}).
The 3-pixel $R$-band check does not alter the test
outcome; we adopt the 5-pixel results for both bands.
No significant intraday variability is detected at
the achieved sensitivity. Measurement precision and
host-galaxy dilution limit sensitivity to nuclear
variations.

\begin{table}
\centering
\caption{IAC80 power-enhanced $F$-tests.}
\label{tab:ic1495_idv_results}
\footnotesize
\setlength{\tabcolsep}{4pt}
\renewcommand{\arraystretch}{1.12}
\begin{tabular}{lccccc}
\hline
Date & Band & $r_{\rm ap}$ & $F_{\rm enh}$ & $F_{\rm crit}$ & $p$ \\
& & (pix) & & & \\
\hline
2025-10-13 & $R$ & 5 & 0.468 & 2.238 & 0.974 \\
2025-10-13 & $R$ & 3 & 0.887 & 2.278 & 0.610 \\
2025-10-14 & $V$ & 5 & 0.993 & 1.985 & 0.495 \\
\hline
\end{tabular}
\par\smallskip
\begin{minipage}{\columnwidth}
\footnotesize
\raggedright
\textit{Notes.} The 5-pixel results are adopted for both
bands; the 3-pixel $R$-band result tests aperture
dependence. $F_{\rm crit}$ is the critical value at
$\alpha=0.01$. Aperture radii are specified in
Appendix~\ref{app:iac80_data}.
\end{minipage}
\end{table}

\subsection{Host structure and global H\,I}
\label{app:host_hi_results}

\subsubsection{Stellar structure}
\label{app:host_results_robustness}

\begin{figure}
\centering
\includegraphics[width=\columnwidth]{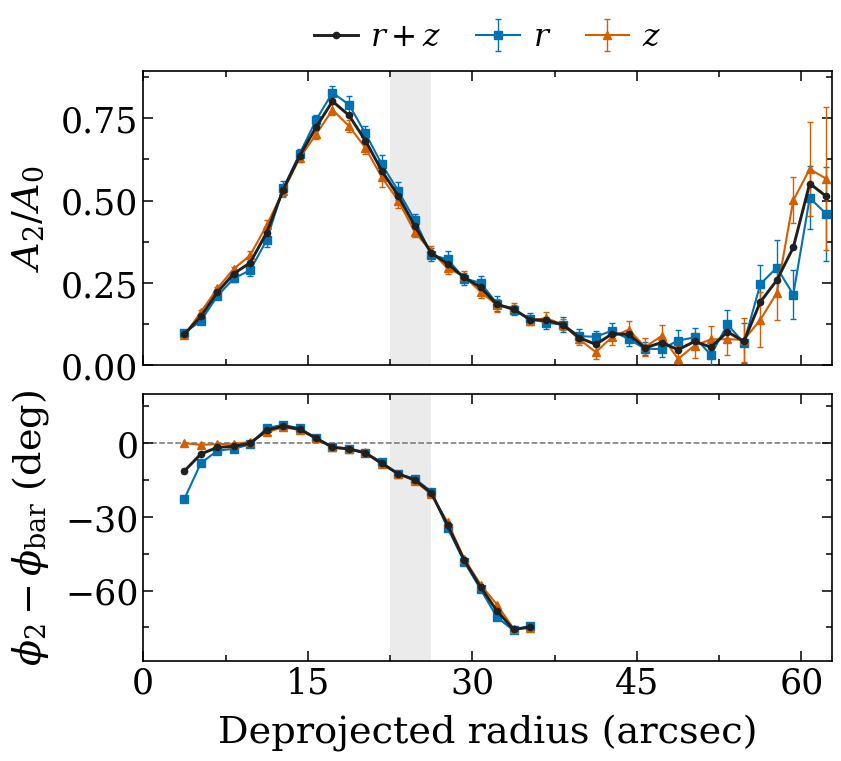}
\caption{Fourier bar diagnostics. The upper panel shows the
deprojected $m=2$ amplitudes in $r,z$ and their combination.
The lower panel shows the phase within $R\leq36^{\prime\prime}$
relative to $\phi_{\rm bar}=53.3^\circ$,
unwrapped for the $180^\circ$ periodicity. Shading marks the
bar-radius bracket.}
\label{fig:ic1495_bar}
\end{figure}

Outer isophotes converge beyond about $40.2^{\prime\prime}$.
The five-component M4 Ferrers radius falls within the Fourier
bracket (Fig.~\ref{fig:ic1495_bar}).
Raising its upper bound from 35 to $55^{\prime\prime}$ changes
the radius by less than $0.002^{\prime\prime}$ and BIC by 0.06.
Other starts give larger radii with poorer fits.
The Fourier bracket is adopted given the fit boundaries and
residual spiral structure.
The $m=1$ signal persists under sector omission and mask and
background reconstruction. Table~\ref{tab:ic1495_host_robustness}
gives the amplitudes, coherence, and scrambling tests.

\begin{table}
\centering
\caption{Supplementary host measurements.}
\label{tab:ic1495_host_robustness}
\footnotesize
\setlength{\tabcolsep}{3pt}
\renewcommand{\arraystretch}{1.08}
\begin{tabular}{@{}p{0.43\columnwidth}p{0.53\columnwidth}@{}}
\hline
Quantity & Result \\
\hline
Integrated $(g,r,z)$ & $(13.707,13.038,12.542)$ AB mag \\
Post-bar scale length & $9.47^{\prime\prime}$ (4.06 kpc) \\
Masked-pixel estimator difference & $<1.4\times10^{-4}$ mag \\
Extrapolated outer light & 0.7--0.8 per cent \\
Maximum alternative $M/L$ shift & 0.061 dex \\
Stellar-mass shift after central excision &
$-0.084$ dex ($R<3^{\prime\prime}$) \\
Projected bar PA & $44.1\pm1.3^\circ$ \\
Deprojected Fourier bar radius &
$22.50$--$26.25^{\prime\prime}$ (9.65--11.26 kpc) \\
M4 Ferrers radius, projected / deprojected &
$20.72^{\prime\prime}$ / $25.17^{\prime\prime}$ (10.79 kpc deprojected) \\
M4 parameters at bounds &
$q_{\rm disc}=0.45$, $c_0=-0.5$, $\sigma_r=10^{\prime\prime}$ \\
Fiducial outer $A_1/A_0$, phase resultant, phase scatter &
$\simeq0.199$, $0.993$, $\sim4.7^\circ$ \\
Weakest jackknife $A_1/A_0$ & $\sim0.11$ \\
Reconstructed $A_1/A_0$, phase resultant & 0.192, 0.994 \\
Scrambling of reconstructed profile &
$k=0$ in $N=5000$, $p_{\rm MC}=1/5001$ for amplitude, phase,
and joint statistic (equation~\ref{eq:finite_mc_probability}) \\
\hline
\end{tabular}
\par\smallskip
\begin{minipage}{\columnwidth}
\footnotesize
\raggedright
\textit{Notes.} The $m=1$ reconstruction uses the same images
with reconstructed masks and backgrounds and a different
random seed.
\end{minipage}
\end{table}

\subsubsection{H\,I profile and dynamical check}
\label{app:hi_results_robustness}

The remeasured linewidth agrees with EDD. The BTFR offset is
consistent with zero within propagated uncertainties
(Table~\ref{tab:ic1495_hi_robustness}).
The velocity estimate uses the unresolved linewidth proxy.
Beam-confusion results are given in Appendix~\ref{app:hi_confusion}.

\begin{table}
\centering
\caption{H\,I profile and linewidth checks.}
\label{tab:ic1495_hi_robustness}
\footnotesize
\setlength{\tabcolsep}{3pt}
\renewcommand{\arraystretch}{1.12}
\begin{tabular}{@{}p{0.41\columnwidth}p{0.51\columnwidth}@{}}
\hline
Quantity & Result \\
\hline
Direct $W_{mx}$ & $293^{+13}_{-18}$~km~s$^{-1}$ \\
Adopted EDD $W_{mx}$ & $304\pm18$~km~s$^{-1}$ \\
$p_{\rm sym}$ range & 0.134--0.246 \\
$M_{\rm bar,atomic}$ & $7.36^{+2.51}_{-1.84}\times10^{10}\,M_\odot$ \\
$f_{\rm atomic,bar}$ & $0.127^{+0.044}_{-0.034}$ \\
$V_{\rm rot,W}$ & $210.3^{+13.3}_{-13.1}$~km~s$^{-1}$ \\
BTFR $\Delta\log M_{\rm bar}$ & $-0.092^{+0.179}_{-0.173}$ dex \\
\hline
\end{tabular}
\par\smallskip
\begin{minipage}{\columnwidth}
\footnotesize
\raggedright
\textit{Notes.} The $p_{\rm sym}$ range spans the 16th--84th
percentiles across prescriptions.
The atomic baryonic fraction is
$f_{\rm atomic,bar}=1.33M_{\rm HI}/M_{\rm bar,atomic}$.
The fiducial asymmetry is in Table~\ref{tab:ic1495_host_summary}.
\end{minipage}
\end{table}

\label{lastpage}
\end{document}